\RequirePackage{amsmath}
\documentclass[smallcondensed,final]{svjour3}     % onecolumn (ditto)

\smartqed  % flush right qed marks, e.g. at end of proof

\usepackage{graphicx}
\usepackage[utf8]{inputenc} % funny chars supported
\usepackage{url}
\usepackage{natbib}
\setcitestyle{authoryear}

\usepackage{IEEEtrantools} % For listing math equations and aligning them

\newcommand{\mr}{\mathrm}
\usepackage{amsfonts}
\usepackage[caption=false]{subfig}
\usepackage{tabularx}
\usepackage{siunitx}
\usepackage{comment}
\usepackage{floatrow}
\newfloatcommand{capbtabbox}{table}[][\FBwidth]

\usepackage{booktabs}
\usepackage{realboxes}
\usepackage[most]{tcolorbox}
\usepackage{todonotes}
\usepackage{enumerate}
\usepackage{colortbl}
\usepackage{makecell}

\newenvironment{ieeeEnumerate}{
    \begin{enumerate}[1)]
}{
    \end{enumerate}
}
\definecolor{Red}{rgb}{0.984313725, 0.807843137, 0.694117647} % Apricot

\definecolor{Orange}{rgb}{1, 0.898039216, 0.705882353} % Peach

\definecolor{Green}{rgb}{0.815686275, 0.941176471, 0.752941176)} % Tea green

\newcommand{\innerQuote}[1]{\lq{#1}\rq{}}
\newcommand{\simpleQuote}[1]{``#1''}
\newcommand{\interviewQuote}[1]{\textit{\simpleQuote{#1}}}
\newcommand{\quoteSkip}{[\ldots]}
\newcommand{\figRef}[1]{Figure~\ref{#1}}
\newcommand{\figsRef}[2]{Figures~\ref{#1}--\ref{#2}}
\newcommand{\tabRef}[1]{Table~\ref{#1}}

\newcommand{\secRef}[1]{Section~\ref{#1}}

\newcommand{\percent}[1]{#1~\%}

\newcommand{\researchObjective}{to investigate the relationship between TD and affective state from the point of view of software practitioners}

\newcommand{\psychologicalRebound}{\emph{psychological rebound}}
\newcommand{\psychologicalRebounds}{\psychologicalRebound{}\emph{s}}

\newcommand{\theme}[1]{\emph{#1}}
\newcommand{\undergoingTD}{\theme{Undergoing TD}}
\newcommand{\forecastingTD}{\theme{Forecasting TD}}

\newcommand{\subtheme}[1]{\emph{#1}}
\newcommand{\procrastination}{\subtheme{Procrastination}}
\newcommand{\elitism}{\subtheme{Elitism}}
\newcommand{\compensation}{\subtheme{Compensation}}
\newcommand{\apprehension}{\subtheme{Apprehension}}
\newcommand{\indeterminable}{\subtheme{Indeterminable}}

\newcommand{\simpleBox}[1]{
    \begin{tcolorbox}[enhanced, arc=3mm, drop shadow]
        #1
    \end{tcolorbox}
}

\newcommand{\RQBox}[1]{\simpleBox{#1}}

\begin{document}

\title{Measuring affective states from technical debt}
\subtitle{A psychoempirical software engineering experiment}

%\titlerunning{Short form of title}        % if too long for running head

\author{Jesper Olsson         \and
        Erik Risfelt \and
        Terese Besker \and
        Antonio Martini \and
        Richard Torkar
        }

\authorrunning{Olsson et al.} % if too long for running head

\institute{J.~Olsson, E.~Risfelt, T.~Besker \at
              Chalmers and University of Gothenburg \\
              Dept.~of Computer Science and Engineering\\
              SE-412 96 Göteborg, Sweden\\
              \email{research@jesperolsson.se}
           \and
           R.~Torkar \at
              Chalmers and University of Gothenburg \\
              Dept.~of Computer Science and Engineering\\
              SE-412 96 Göteborg, Sweden \at
              Stellenbosch Institute for Advanced Study (STIAS)\\
              Wallenberg Research Centre at Stellenbosch University\\ Stellenbosch, South Africa\\
           \and
           A. Martini \at
              University of Oslo\\
              Dept.~of Informatics\\
              N-0373 Oslo, Norway\\
}

\date{Received: date / Accepted: date}
% The correct dates will be entered by the editor

\maketitle

\begin{abstract}
  Software engineering is a human activity. Despite this, human aspects are under-represented in technical debt research, perhaps because they are challenging to evaluate.

This study's objective was to investigate the relationship between technical debt and affective states (feelings, emotions, and moods) from software practitioners. Forty participants ($N=40$) from twelve companies took part in a mixed-methods approach, consisting of a repeated-measures ($r=5$) experiment ($n=200$), a survey, and semi-structured interviews.

The statistical analysis shows that different design smells (strong indicators of technical debt) negatively or positively impact affective states. From the qualitative data, it is clear that technical debt activates a substantial portion of the emotional spectrum and is psychologically taxing. Further, the practitioners' reactions to technical debt appear to fall in different levels of maturity. 

We argue that human aspects in technical debt are important factors to consider, as they may result in, e.g., procrastination, apprehension, and burnout.
\end{abstract}

\section{Introduction}\label{sec:introduction}
Software engineering is very much a human activity, but this is sometimes forgotten. When proposing hypotheses, analyzing results, and discussing implications for the industry, we researchers sometimes neglect to factor in human aspects~\citep{Lenberg2015}. So, too, is the case for technical debt research (except for a handful of studies on morale, e.g.,~\citep{morale-TD}). This paper intends to amend this deficit and provide evidence showing that technical debt has noticeable adverse effects on software practitioners' feelings.

Technical Debt (TD) is a financial metaphor~\citep{wycash}, typically used within software engineering to explain long-term costs of short-term benefits~\citep{ampatzoglou2015financial}. It is a communicative aid for bridging the knowledge gap between software practitioners and business decision makers. Hence, if the metaphor was to miscount (or not account for) pivotal cost-benefit factors, the effect could be detrimental to software companies.

The current definition of TD was agreed upon during the 16162 Dagstuhl seminar~\citep{Dagstuhl}: \simpleQuote{In software-intensive systems, technical debt is a collection of design or implementation constructs that are expedient in the short term, but set up a technical context that can make future changes more costly or impossible. Technical debt presents an actual or contingent liability whose impact is limited to internal system qualities, primarily maintainability and evolvability.}

The definition is nuanced, incorporates decades of research, and offers a shared understanding of TD\@. Among many other things, it emphasizes that TD is a software development artifact in its own right and that TD acquisition is not necessarily intentional nor visible. A list of various consequences was also synthesized, but it fell short in recognizing the effects of TD on the human aspects of software engineering.

This paper aims to fill that gap by assessing five different design smells (proxies for design TD) to understand if, how, and why these smells impact participants' affective states during their development work.

In this study, we address this gap by employing a mixed-methods approach (including an experiment) and following guidelines for psychoempirical software engineering research (\simpleQuote{research in software engineering with proper theory and measurement from psychology}~\citep{psychoempirical-SE}). The study collected empirical data ($n=200$ data points from $N=40$ participants) on how design TD influences the so-called affective state of software practitioners. Applying Bayesian multi-level models revealed, among other findings, strong evidence that certain design smells (notably cyclic-dependencies) caused the subjects displeasure. The qualitative analysis suggests that many practitioners experience anxiety from high amounts of TD, and their responses vary along a maturity scale.

In more concrete terms, the research objective of this study is \researchObjective{}. This objective is supported by three research questions, which are listed below and further elaborated on in \secRef{sec:methodology}.

\RQBox{
    \textbf{RQ1:} How do software practitioners' affective state change in the presence of design smells?
    
    \bigskip
    
    \textbf{RQ2:} How do changes in affective state align with professional characteristics (e.g., formal education, work experience, or work context)?
        
    \bigskip
    
    \textbf{RQ3:} How do software practitioners reason about the relationship between affective states and technical debt?
}

The results of this study provide important insights and show that psychological factors also need to be acknowledged as a consequence of TD\@. The results show, for instance, that different kinds of design smells impact participants' affective states differently. When assessing how the affective state aligns with the practitioners' professional characteristics, the results show %for instance that these different states most likely correlate with the practitioners' work experience.
that work experience correlates with submissiveness. Lastly, practitioners reason, e.g., that negative affects often coincide with TD, but can be viewed as opportunities for improving the code base.

The sections of this paper are laid out as follows. The following section presents related work in the research areas of TD and human aspects of software engineering, individually and jointly. \secRef{sec:methodology} describes the research design and methods employed. Next, Sections~\ref{sec:quantitative}--\ref{sec:qualitative} present the quantitative and qualitative analyses, respectively. The study is discussed in \secRef{sec:discussion}, limitations and threats to validity are presented in \secRef{sec:threats}, and the paper is concluded in \secRef{sec:conclusion}.

\section{Related Work}\label{sec:related_work}
Much of the current literature on Technical Debt (TD) pays particular attention to technical or financial perspectives. This study breaks with such traditions to observe TD through the lens of human aspects of software engineering. Hence, for full appreciation, the reader should be familiar with the background of the two research fields.

Recounted firstly is previous research on TD in general. Appropriate nomenclature and central findings are outlined before introducing the specific type of TD investigated in this study. Secondly, we describe software engineering research on human behavior, emphasizing recent studies on the topic of feelings, emotions, and moods, and the recommendations concerning measurement instruments from psychology. One of those instruments, the Self-Assessment Manikin (SAM), was employed in this study and is explained in detail.

Once these two branches (i.e., the research area to be broadened, and the facet used to do so) have been covered, related work is listed. That is, existing research items that have used similar lenses and investigated challenges encountered in the TD literature. Those items are briefly reviewed to clarify how this study fits into the current body of knowledge.

\subsection{Previous Research on Technical Debt}

Technical Debt (TD) was conceptualized a few decades ago by~\citet{wycash} as a financial metaphor for how early misunderstandings of a problem domain might hamper future development unless the software is refactored to incorporate knowledge gained. Since then, the term has received much attention in both academia and industry. Today, the metaphor is widely used as a communicative aid for explaining internal software quality problems to non-technical stakeholders by emphasizing the extent to which the software must compromise its ability to meet the needs of the future to meet the needs of the present~\citep{wycash,Dagstuhl,ampatzoglou2015financial,fernandez2017identification,ernst2015measure}.

%The concept of TD was redefined during the Dagstuhl research seminar of 2016 (the ISO 16162 definition of technical debt) to establish a common language that encompasses the recent scientific findings~\citep{Dagstuhl}. The definition emphasizes that TD is not only a mechanism for communication but also a software development artifact in its own right. Further,~\citet{Dagstuhl} argue that TD acquisition is not necessarily intentional or visible, and the discovery thereof may catch stakeholders by surprise.

One of the main strengths of TD is that much of its terminology originates from finance. As noted by~\citet{ampatzoglou2015financial}, the two most commonly used terms in TD research are \textit{principal} and \textit{interest}, i.e., the cornerstones of financial debt. In software engineering, the former expresses the effort required to turn the current quality of some development artifact into its optimal level---the latter concerns how this sub-optimal level of quality leads to extra effort in later development iterations.

Several studies have shown that TD has significant negative consequences that can be detrimental to software companies~\citep{exploration-TD,li2015systematic,besker-SLR,ampatzoglou2015financial,fernandez2017identification}. The interest does away with a substantial portion of development time~\citep{cost-TD,besker2019software}, and may grow non-linearly if left unattended~\citep{martini2017interest}. Further, TD tracking and TD management are uncommon in the software industry, and when encountered, the processes are typically immature~\citep{guo2011tracking,ernst2015measure,martini2018technical}.

Despite its severity, TD is difficult or impossible to measure directly, and assessments typically rely on measurement proxies known as software smells, i.e., indicators of (internal) software quality issues~\citep{fontana2017arcan,ganesh2013towards,garcia2009toward,sharma2018survey}. Naturally, empirical studies, such as this one, face the same issue when they need to exemplify TD items.

So far, we have outlined the previous research on TD in general, by giving an account of its history, terminology, and critical findings. The next paragraphs will focus on a type of TD known as Design TD (DTD), which our investigation is based on. 

True to its name, DTD is TD found in software design, i.e., sub-optimal constructs in the software system's structure and behavior. As such, its boundary to, e.g., architectural TD (ATD), is disputed. Some researchers merge the two~\citep{exploration-TD}. Others separate them~\citep{li2015systematic,alves2016identification} according to definitions that typically are too vague or subjective to form disjunct sets~\citep{alves2014towards,alves2016identification}.

Such disagreements propagate to the categorization of software smells~\citep{garcia2009toward}, which results in some smells, e.g., cyclic dependencies and hub-like dependencies being considered either design smells~\citep{ganesh2013towards} or architectural smells~\citep{fontana2017arcan}.

To reduce the risk of misinterpretation, this study will not merge the two categories. The investigation is concerned with small, local problems, in isolated parts of the software system that can be comprehended easily. The findings should not be confused with the large concerns highlighted in recent ATD research, e.g.,~\citet{ernst2015measure,besker-SLR}.

\subsection{Previous Research on Human Aspects of Software Engineering}

A growing body of literature recognizes the importance of interdisciplinary research between software engineering and psychology~\citep{cruz2015forty}. Both academia and the industry acknowledge that software engineering tasks are human activities and, thus, impacted by human aspects~\citep{boehm1988understanding,feldt2010links,colomo2010study,tamburri2013social,fagerholm2015performance}.

For many years, such studies were dispersed, but in 2015 Behavioral Software Engineering (BSE) was proposed as a common platform for research concerned with \simpleQuote{the study of cognitive, behavioral, and social aspects of software engineering performed by individuals, groups, or organizations}~\citep{Lenberg2015}.

Out of the many tracks in this research area, one concerns \emph{affective states} (or \emph{affects}, for short), i.e., feelings, emotions, and moods. Previous studies have linked affects to, e.g., debugging performance~\citep{khan2011moods}, analytical ability~\citep{happy-analytical}, and productivity~\citep{do-feelings-matter}.

This study is placed firmly within this track and is part of a sub-field called psychoempirical software engineering (PSE), i.e., software engineering studies that use theory and measurements from psychology~\citep{psychoempirical-SE}. This article follows the \citet{psychoempirical-SE} guidelines for conducting PSE research.

According to these guidelines, this study's objective is best met by subscribing to the \textit{dimensional framework} and employing the \textit{Self-Assessment Manikin} (SAM) instrument for measuring affective states~\citep{psychoempirical-SE}. Within the dimensional framework, affects are expressed through several distinctive dimensions, e.g., the models represent affective states along three continua: pleasure--displeasure (valence), arousal--nonarousal (arousal), and dominance--submissiveness (dominance) \citep{psychoempirical-SE,Russell1977ThreeDimensions}.

In more concrete terms, according to~\citet{do-feelings-matter}, these dimensions can be understood as follows. Valence is the attractiveness (or adverseness) of an event, object, or situation, while arousal is the intensity of emotional activation or the sensation of being mentally awake and reactive to stimuli. Finally, dominance is the sensation of control of the situation; one's skills are perceived to be higher than the challenge level for the task.

\begin{figure}
    \centering
    \includegraphics[width=1\linewidth]{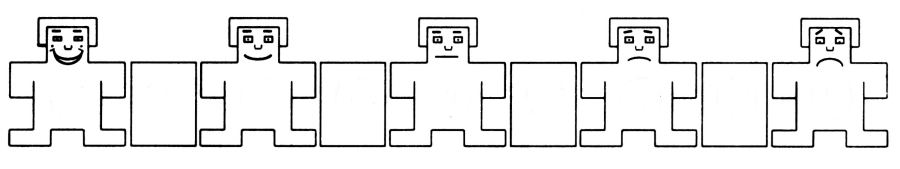}
    \includegraphics[width=1\linewidth]{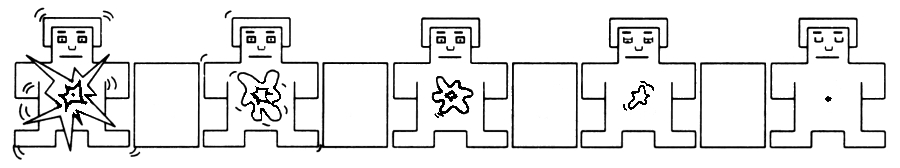}
    \includegraphics[width=1\linewidth]{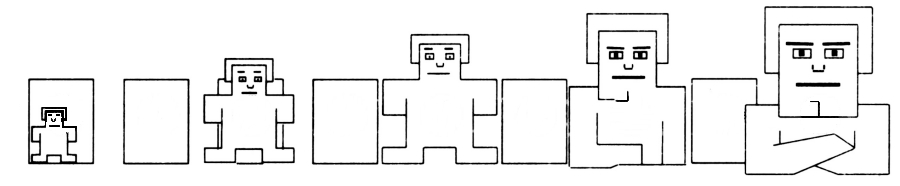}
    \caption{The SAM measurement instrument. SELF ASSESSMENT MANIKIN \copyright{} Peter J.~Lang 1994}
    \label{fig:sam}
\end{figure}

The recommended instrument, the SAM, measures affects through pictorial representations (\figRef{fig:sam}) of the three dimensions of the models~\citep{psychoempirical-SE,SAM-first,SAM-original,power-of-affect}. Developed by~\citet{SAM-first}, the instrument has, over the decades, been subjected to extensive validation research~\citep{SAMValidity} and seen used in numerous studies, see~\citep{SAMValidity,SAM-critic}.

According to~\citet{SAM-original}, the graphic design of the SAM has many benefits. The lack of verbal components means that the SAM can be administered to a broader population range, including individuals with a non-English mother tongue or language disorders, and children. Additionally, the SAM can measure direct affective reactions, as it can be filled out in a short amount of time and eliminates cognitive processing~\citep{power-of-affect}. Further,~\citet{SAMValidity} argues that the use of stylized characters, as opposed to photographs of humans, makes the SAM less susceptible to many types of biases.

However, because SAM relies on self-reporting, the scores are not standardized according to objective reference points. Although individuals are consistent with themselves (within measurement), the ratings cannot be assumed to be consistent between individuals (between measurement)~\citep{psychoempirical-SE}. In other words, two individuals could rate the same affective state in two different ways. Consequently, investigations administrating the SAM should follow a within-subject (or repeated measures) design~\citep{psychoempirical-SE}, which also follows the latest recommendations in general (see~\citep{gelman18nhst}).

Additionally, it is important to recognize that the SAM is not suited for all types of affective state research;~\citet{psychoempirical-SE} emphasize that the instrument is designed to measure affective reactions in response to a stimulus (in our case, design smells). For example, the SAM would be unfit for studies aiming to investigate how happy software practitioners are, generally~\citep{psychoempirical-SE}.

The SAM is protected by copyright law, but the instrument and instructions for proper administration~\citep{SAM-protocol} are available for non-profit academic research\footnote{Information about how to obtain the SAM can be found at \url{https://csea.phhp.ufl.edu/Media.html}}.

\subsection{Interdisciplinary Research on TD and Human Aspects \label{sec:TD-and-human-aspects}}

Data from several secondary studies reveal that few TD studies have investigated the relationship between TD and human aspects~\citep{exploration-TD,li2015systematic,ampatzoglou2015financial,alves2016identification,fernandez2017identification,besker-SLR}. Rather, the predominant concerns have been technical and financial aspects, e.g., software quality or cost of future changes.

When human aspects are addressed in TD research, the most frequently investigated topic is morale. A negative correlation was proposed early by~\citet{exploration-TD} based on anecdotal evidence found in web blogs. Since then, empirical investigations have corroborated the connection, including previous articles of our own, see~\citep{morale-TD}.

\citet{td-folklore} performed a survey on TD folklore and found medium to high consensus among software practitioners that TD is related to their morale. In conjunction with interviews, a survey was also carried out by~\citet{morale-TD} to determine how occurrence and management of TD affect developers' morale. Their findings show that the existence of TD negatively impacts morale, but also that morale is increased by proper TD management.

Although a common misconception, morale is not the same thing as affective state~\citep{graziotin2015affect, peterson2008group}. Hence, to the best of our knowledge, there are no previous TD studies investigating affects and even fewer that directly measure how software practitioners respond to TD items.

In addition to morale, some empirical studies have offered evidence for TD harming the software practitioner's psychology. \citet{lim2012balancing} found that developers are more reluctant to incur TD because its consequences become a part of their daily work. Similarly, such reluctance may arise due to developers predicting that the sub-optimal construct needs to be corrected sooner or later, and that task would fall on them~\citep{yli2014sources}. However, these findings were somewhat opportunistic and limited, as neither study set out with the research objective of investigating such questions.

TD research has thus far shown lukewarm interest in the relationship between TD and human aspects. However, the topic has also been approached from the PSE direction, and those studies present interesting empirical findings. \citet{graziotin2017unhappiness} surveyed software practitioners concerning causes for unhappiness, and established that low code quality and coding practices, and being stuck in problem-solving, were among the most significant factors. Additionally, in a later study, \citet{graziotin2018happens} investigated the adverse effects of developer displeasure and found, among many other types of consequences, \emph{lower code quality} and \emph{discharging code} (extreme cases of productivity and quality drop, in the form of deleting parts of the code base).

Not only are these factors intimately connected with TD, but they pose the threat of vicious cycles: Low code quality causes unhappy developers, and unhappy developers produce low-quality code. Unfortunately, the studies did not drill down into this problem, which could answer questions such as its probability and severity. Nor was the issue approached specifically from the TD perspective. Clearly, our study differs from the previous PSE studies, as it seeks to investigate affects regarding specific TD items.

In conclusion, prior research shows that investigating human aspects concerning TD is a promising prospect. To manage TD more effectively, we need to understand how software practitioners, as human beings, can be factored into the trade-offs between short-term and long-term benefits. However, the current body of knowledge is limited, and both academia and the software industry would likely benefit from further clarification.

\section{Methodology \label{sec:methodology}}
As suggested in the previous section, our research topic has received little attention despite interesting initial findings. Consequently, the study design must acknowledge the limitations posed by such research gaps, e.g., validation against previous findings may be impossible.

One of the countermeasures implemented in our design is choosing a mixed-methods approach, i.e., collecting both quantitative and qualitative data. This decision is appropriate because it enables the study to improve validity, e.g., the results from one analysis could corroborate or rebut findings from the other. In this study, data were gathered from three sources: A repeated-measures experiment (quantitative), a questionnaire (quantitative), and a semi-structured interview (qualitative).

Another central countermeasure is the high transparency achieved by providing a replication package for this publication\footnote{\url{http://doi.org/10.5281/zenodo.4537801}}. It contains complementary information and all material needed for reproducing the study, as it is infeasible to present all details within the scope of this article.

\begin{figure}
    \centering
    \includegraphics[width=0.8\linewidth]{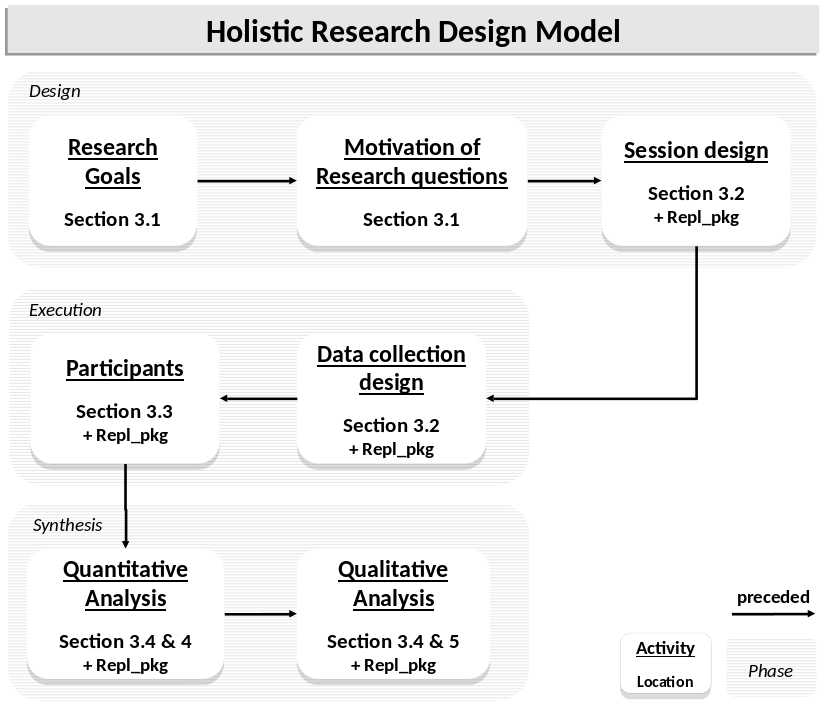}
    \caption{The design of the study.}
    \label{fig:res-des}
\end{figure}

To demonstrate this study's overall study design, we have constructed a holistic research design model as illustrated in \figRef{fig:res-des}. As shown, this study was conducted in three different phases: a design, an execution, and a synthesis phase. The figure also illustrates the different performed activities within each phase and references the sections describing these activities. If more information exists in the replication package, this is also pointed out (using the tag repl\_pkg).

\subsection{Goals}

This study seeks to examine the relationship between design smells and software practitioners' affective states. Thus, it tries to understand the importance of human aspects as a factor in TD\@. Among other things, we hope that the answers to our research questions will spark further interest in considering software practitioners when making trade-offs between short-term and long-term benefits. This goal begs for persuasive evidence, which can be provided through empirical research.

RQ1 (see \secRef{sec:introduction}) will be answered by conducting a within-subjects experiment. The data are analyzed via (Bayesian) statistical analysis: We employ dynamic Hamiltonian Monte Carlo to sample multi-level models. This research question aims to investigate the actual relationship between affects and DTD, without being colored by the participants' (nor the researchers') preconceived notions. As for delimitations, this RQ will examine a handful of design smells and consider affects from the presented models' perspective alone.

The motivation behind RQ2 is to see what role individual differences play. Because the study examines affects, the experimental units must be human participants, which opens up many exciting characteristics that could be studied. However, while data for various factors could be collected with ease, there are trade-offs to consider, e.g., transparency and confidentiality. Since the data are open (see the replication package), many characteristics that could easily identify an individual (e.g., gender or ethnicity) were not recorded.

Finally, RQ3 was included to understand the topic's appearance in the software industry. Hence, this research question is broader than the other two and of a more exploratory nature. Giving voice to the practitioners' reflections on affects and TD can increase understanding in a broader context and reveal peripheral issues.

\subsection{Session Design}

\begin{figure}
    \centering
    \includegraphics[width=\textwidth]{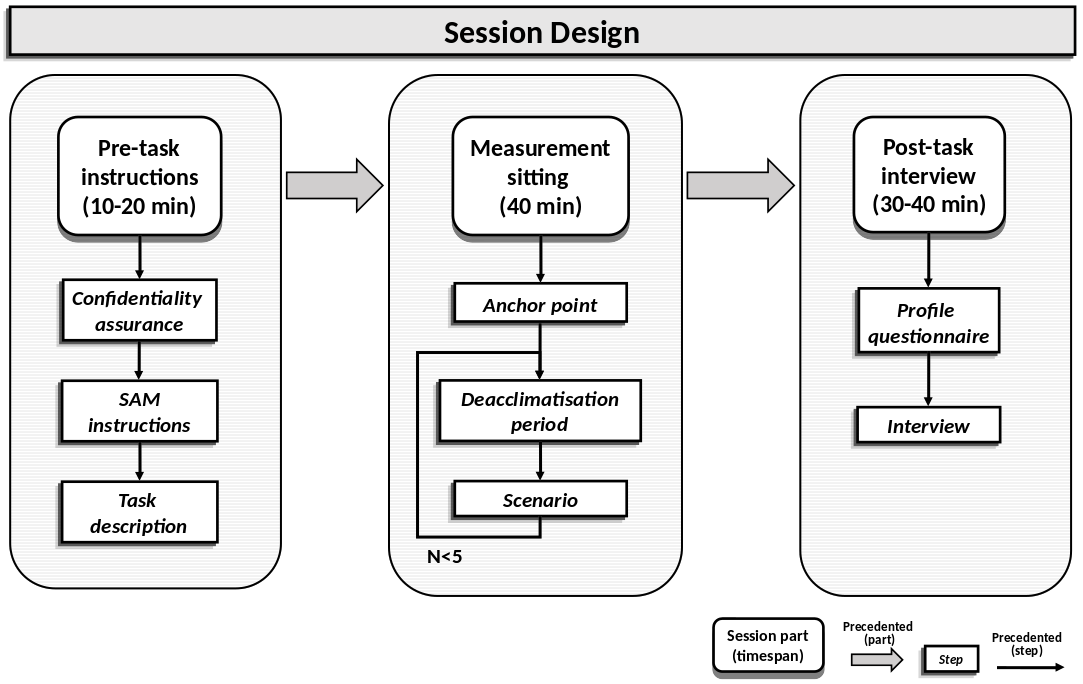}
    \caption{The session view of the study: $90$-minutes sessions, conceptually comprising three parts with eight steps.}
    \label{fig:experiment-design}
\end{figure}

As this study collected three sets of data, its design is a substantial part of this article. Since there are many constructs to keep track of and clarify, we will use a few different viewpoints. The first viewpoint is that of \textit{sessions} and is modeled in \figRef{fig:experiment-design}.

From this perspective, the study was designed as $90$-minute sessions, one for each participant. At the start of their session, the participant received instructions (pre-task instructions) outlining the study and the session. The participant obtained these in three steps:

\begin{ieeeEnumerate}
    \item reading, understanding, and signing a document describing the treatment of, and their rights regarding, collected data (confidentiality assurance);
    \item listening to instructions for, and seeing examples of, how to use the measurement instrument---which relies on self-reporting (SAM instructions); and
    \item hearing a description of what activities they will perform during the experiment (task description).
\end{ieeeEnumerate}

Next, during the second part of the session (measurement sitting), quantitative data were collected from a repeated-measures experiment. For this part, as well, the participant went through three steps (please note that being of a repeated-measures design, the second and third steps were conducted five times):

\begin{ieeeEnumerate}
    \item using the measurement instrument on a practice task (anchor point);
    \item pausing briefly (deacclimatization period); and
    \item using the measurement instrument on a task (scenario).
\end{ieeeEnumerate}

In the last part (post-task interview), the two remaining data sets were gathered: quantitative data from a questionnaire and qualitative data from a semi-structured interview. These were presented to the participant in one step each:

\begin{ieeeEnumerate}
    \item filling out answers to questions about their professional experience with software (profile questionnaire) and;
    \item talking and answering questions about how they perceived the study and their view of code maintainability and feelings (interview).
\end{ieeeEnumerate}

Thus, the session perspective is concluded. This description has given an overview of what the participants did, between being greeted by the researchers to saying goodbye. It also introduced concepts that are key to understanding the study design but did so on a high abstraction level. Further details on these concepts can be found in the replication package.

\begin{figure}
    \centering
    \includegraphics[width=\textwidth]{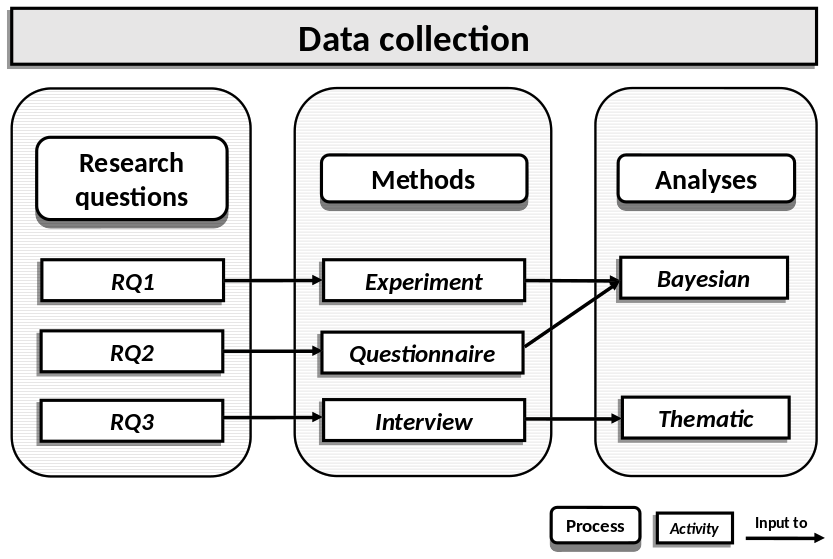}
    \caption{The relationships between the RQs, methods, and analyses.}
    \label{fig:rq-method-analysis}
\end{figure}

Next, we consider the study from the perspective of \textit{data collection}. Three sources of empirical data (experiment, questionnaire, and interview) were gathered from the participants. As shown in \figRef{fig:rq-method-analysis}, each of these data sets was designed around one of the RQs, i.e., the experiment for RQ1, the questionnaire for RQ2, and the interview for RQ3\@. Similarly, the experiment data and the questionnaire were modeled in the same statistical analysis, while the interview data underwent thematic analysis.

First, the experiment set out to understand the relationship between affects and DTD\@. From this goal, it followed that, ideally, all factors except for the amount of design debt (explanatory variable), should remain constant. Then, what was measured was the participants' affective state in terms of valence, arousal, and dominance (response variables).

However, since the experiment was of the repeated-measures variety, its design was more complicated. While the explanatory variable still represented the amount of design debt, there was not one but five such variables (one for each repetition or \emph{scenario}). In other words, as the participant progressed through the experiment, they would encounter five different scenarios: ScA, ScB, ScC, ScD, and ScE\@. Within each scenario, the participant received one treatment and then reported their affective state.

Because design debt is difficult to measure, each response variable had two levels and represented whether its design smell (see \tabRef{tab:final-scenarios}) was present or had been refactored away. That is, the scenario variant where the smell had been removed had a lower ($L$) amount of technical debt than its partner variant ($H$).

\begin{table}
    \centering
    \caption{The scenarios used in the experiment and the smells they embody.}
    \begin{tabularx}{\textwidth}{lXX}
        \toprule
        ID  & Smell & Smell category \\ 
        \midrule
        ScA & Missing Encapsulation & Encapsulation smell \\
        ScB & Missing Hierarchy & Hierarchy smell \\
        ScC & Broken Modularization & Modularization smell \\
        ScD & Cyclically-Dependent Modularization & Modularization smell \\
        ScE & Rebellious Hierarchy & Hierarchy smell \\
        \bottomrule
    \end{tabularx} 
    \label{tab:final-scenarios}
\end{table}

The scenarios were derived from~\citet{refactoring-design-smells}, which in turn is based on~\citet{ganesh2013towards}. Because smells are not necessarily indicative of definite quality problems~\citep{sharma2018survey}, smell catalogs such as~\citet{garcia2009toward} were considered inappropriate for the experiment.

Moving on to the second method, the questionnaire aimed to investigate how professional characteristics factor into the participants' responses. The questions are listed in \tabRef{tab:questionnaire-questions}.

\begin{table}
    \centering
    \caption{The questionnaire.}
        \begin{tabularx}{\textwidth}{llX}
        \toprule
        ID & Type & Description \\ 
        \midrule
        Q1 & Closed & My highest level of completed academic education is \rule{1cm}{0.4pt} \\
        Q2 & Closed & My education major (e.g., computer science, electrical engineering, software engineering, \ldots) was \rule{1cm}{0.4pt} \\
        Q3 & Closed & I have working experience with software for \rule{1cm}{0.4pt} years. \\
        Q4 & Closed & My current role (e.g., architect, developer, tester, \ldots) is \rule{1cm}{0.4pt} \\
        Q5 & Closed & The programming language I am most experienced in is \rule{1cm}{0.4pt} \\
        Q6 & Closed & My currently preferred programming language is \rule{1cm}{0.4pt} \\
        Q7 & Closed & Most of my working experience comes from the following domain (e.g., telecom, healthcare, finance, \ldots) \rule{1cm}{0.4pt} \\
        Q8 & Open & Do you have any additional comments concerning this questionnaire? \\
        \bottomrule
    \end{tabularx} 
    \label{tab:questionnaire-questions}
\end{table}

The third method, the interview, was designed to answer RQ3 and explore the topic of TD and human aspects beyond the delimitation of this study. Because the quantity of previous studies is limited, the study gains extra benefits from validating and contextualizing its findings. Hence, caution should be exercised when limiting the participants' divergent thinking and, thus, the data's richness. Therefore, the participants were not constrained to talk merely about DTD\@.

Instead, the participants were allowed to speak more or less freely about their perception of affects and software maintainability. The questions listed in \tabRef{tab:interview-questions} were asked at opportune times during the interview to light-handedly steer it. These were complemented by probing questions, i.e., follow-up questions to the participants' reasoning.

\begin{table}
    \centering
    \caption{The common questions of the semi-structured interview. The thematic analysis used to answer RQ3 is considered a subset (highlighted in green) of the interview questions.}
        \begin{tabularx}{\textwidth}{llX}
        \toprule
        ID & Type & Description \\ 
        \midrule
        IQ1.1 & Open & Could you please tell us more about your daily work. What type of tasks do you normally encounter? \\
        IQ1.2 & Open & How do those tasks make you feel? \\ 
        IQ1.3 & Closed  & Do you face challenges in those tasks?\\
        IQ1.4 & Open  & How do those challenges make you feel?\\
        IQ1.5 & Closed  & Are those feelings frequent? \\
        IQ2 & Open & In contrast to challenging tasks, what sorts of feelings would you say you get from routine tasks? \\
        IQ3 & Closed & Do you think that anything outside of this experiment did impact your responses today? \\
        \rowcolor{Green}
        IQ4.1 & Open & Would you please tell us how you experienced the code examples? \\
        \rowcolor{Green}
        IQ4.2 & Open & What about the software design in the examples? \\
        \rowcolor{Green}
        IQ5 & Open & What would you say are the differences between the scenarios we provided and software one encounters in industry? \\
        IQ6 & Closed & Did you find SAM difficult to use or understand? \\
        IQ7 & Open & That was all of the questions that we had for you. Is there anything you would like to add? \\
        \bottomrule
    \end{tabularx} 
    \label{tab:interview-questions}
\end{table}

Because the interviews had a broader scope than this study, the thematic analysis used to answer RQ3 considered a subset (highlighted in green) of the interview questions, namely IQ4.1, IQ4.2, and IQ5.

Thus, the data perspective is concluded. It presented how the research questions can be traced to the selected methods and analyses. Further, the general structure of the methods was explained, including the questions asked of the participants.

\begin{figure}
    \centering
    \includegraphics[width=\textwidth]{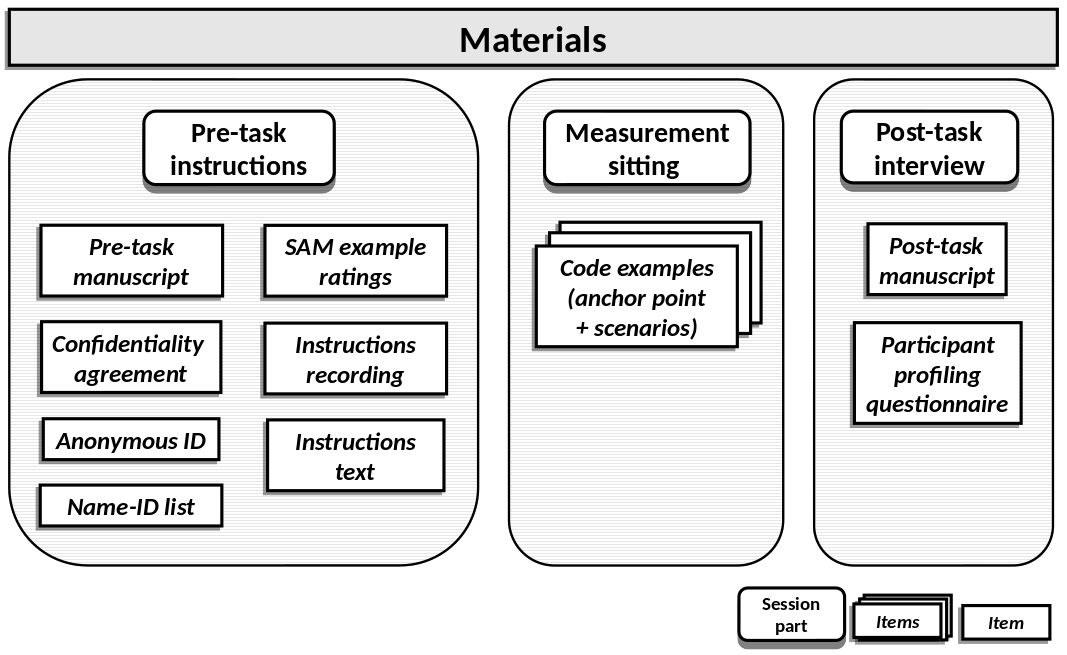}
    \caption{The experimental materials used in the different parts of the session.}
    \label{fig:materials}
\end{figure}

The third and final perspective is the \textit{materials} perspective, which is illustrated in \figRef{fig:materials}. Their description is deferred to the replication package, where the experimental protocol also is included. 

\subsection{Sample}
Forty software practitioners from $12$ companies participated in this study. The participants were obtained through convenience sampling, but covered a diverse set of professional characteristics, e.g., their experience came from many different business domains (such as automotive, finance, and renewable energy) and ranged from $1$ to $35$ years. All participation was voluntary and based on informed consent and anonymity.

\subsection{Analysis Procedure}

Two different analyses were performed in this mixed-methods study. For the quantitative part, a Bayesian statistical model was implemented and executed in \textit{R}~\citep{r}. The procedure is available in the replication package.\footnote{\url{http://doi.org/10.5281/zenodo.4537801}}

The qualitative data was analyzed by following the guidelines for thematic analysis by~\citet{braun2006using}. Thematic analysis is frequently applied in both psychology~\citep{braun2006using} and software engineering~\citep{cruzes2011recommended}. %It is a flexible approach tailored to the research needs. %It is suitable for making the findings accessible to a broader audience, including those with little or no experience of qualitative research~\citep{braun2006using}.

The flexibility of thematic analyses stems from several choices that the researchers must make when deciding how to conduct the analysis (for a discussion about each choice's advantages and disadvantages, see~\citep{braun2006using}). For this study, the analysis was \emph{inductive}, searched for \emph{semantic themes} and theorized \emph{essentialistically}. In other words, we coded the interview transcripts in a data-driven fashion without trying to fit them into a pre-existing coding frame. Themes were then identified and interpreted based on what was explicitly articulated within the data set. %The theorization assumed a simple relationship meaning, experience, and language.

The primary reason for these decisions is the small amount of previous research on the relationship between TD and the human aspects of software engineering. For example, the \emph{inductive} approach does not rely on existing theory to the same extent as the \emph{theoretical}. Similarly, it seemed more prudent to identify the themes at the \emph{semantic} level, given the exploratory nature of this investigation. Otherwise, the likelihood of projecting personal beliefs onto \emph{latent} themes could be excessive. The same reasoning underpinned the choice of performing an \emph{essentialist} analysis. In particular, previous research on human aspects of TD did not seem to lend sufficient support for theorizing socio-cultural contexts and structural conditions (beyond little more than pure speculation), as is sought with the \emph{constructionist} perspective.

Since the qualitative analysis aimed to discover the most central ideas and themes (rather than most, or all of them), the analysis's size was determined by salience rather than (thematic) saturation~\citep{salience}. This decision is somewhat uncommon in software engineering research, so a short motivation is in order. 

Salience is the idea of analyzing qualitative data regarding the most prominent items, and can be contrasted with saturation, i.e., until the set of all unique items is \textit{believed} to have been exhausted. For a broad range of research objectives, saturation would be superfluous, as salient items are, unsurprisingly, more prevalent and more culturally significant than non-salient items~\citep{salience}. In other words, many research questions can be answered with smaller sample sizes than what would be required to claim saturation.

The point at which thematic saturation is reached depends not only on the domain size, but also on the number of responses per person~\citep{salience}. Consequently, salience may be the more appropriate alternative when it is difficult to know the size of the domain or the set of ideas~\citep{salience} (as is the case in this study).

At the same time, the importance of probing questions should not be overlooked: When the investigation aims to obtain most of the most important ideas and themes in a domain (as is frequently the case in qualitative research and particularly in open-ended interviews), a smaller sample with extensive probing is commonly more productive than a large sample with casual or no probing~\citep{salience}. Thus, salience should be used with caution, unless the data collection is designed with this in mind.

Because $10$ interviews are sufficient to reliably capture up to \percent{95} of the most salient ideas~\citep{salience}, that number of data items was randomly selected for the data set (out of the $39$ items in the interview data corpus).\footnote{A single participant asked not to be recorded during the interview and could thus not be included.} Indeed, this study's necessary sample size might be even lower, as we used probing techniques during the interviews, e.g., repeating phrases the interviewee uttered when working with the scenarios and asking for more information.

\begin{comment}
\subsection{Deviations}

\begin{itemize}
    \item Not all participants were interested in modeling the scenarios or were unfamiliar with the practice. When discovered (and after that), the researchers downplayed the task's importance and made it optional.
    \item The researchers intended to not initiate small talk during deacclimatization, to accommodate participants who preferred silence. However, some participants seemed not to dare to initiate conversation, despite showing signs of discomfort. In those instances, the researchers adjusted.
    \item Some participants occasionally slipped into Swedish when struggling to vocalize their thoughts. The researchers allowed this but tried to steer the participant back to English gently. In one instance, all conversation was held in Swedish at the participant's request.
    \item One participant filled out the SAM before the time was due, despite repeated clarifications.
    \item One participant received the scenarios according to an incorrect treatment pattern (HLHLH).
    \item One participant chose to report their level of academic education in the questionnaire falsely. Despite the participant verbally informing that they had not completed their master studies, and despite the researchers pointing out the alternative for \emph{some master studies}, the participant filled in \emph{master degree}.
\end{itemize}
\end{comment}

\section{Quantitative Analysis and Results \label{sec:quantitative}}

Forty subjects participated in the experiment, and each subject contributed with five measurements to estimate our outcomes. Also, the following data were collected: Educational level (e.g., bachelor), the example used (the ten experimental artifacts, i.e., five artifacts in $L$ and $H$ setting), academic major (e.g., computer science), role (e.g., designer), language experience (e.g., Java), entities (i.e., level of complexity of the artifact), and years of work experience. The latter was scaled in order to improve sampling (i.e., $(x_i - \bar{x})/x_{\sigma}$). 

Given the three outcomes valence, arousal, and dominance $\{V, A, D\}$, and the predictors listed above, the data consists of a matrix with $200$ observations (rows) and $11$ variables (columns), with no missing data.\footnote{The dataset, with analysis scripts and a \texttt{Docker} image, can be found at \url{http://doi.org/10.5281/zenodo.4537801}. \texttt{R} 4.0.2, \texttt{rstan} 2.21.2, and \texttt{brms} 2.13.9 was used for the analysis~\citep{r,brms,brms2,rstan}}

% \begin{figure*}
% \centering

% \subfloat[Valence ($V$)]{%
%   \includegraphics[scale=0.25]{figure/plots/hist-V.pdf}%
%   \label{fig:valence}%
% }\qquad
% \subfloat[Arousal ($A$)]{%
%   \includegraphics[scale=0.25]{figure/plots/hist-A.pdf}%
%   \label{fig:arousal}%
% }\qquad
% \subfloat[Dominance ($D$)]{%
%   \includegraphics[scale=0.25]{figure/plots/hist-D.pdf}%
%   \label{fig:dominance}%
% }
% \caption{Histograms of the outcomes $\{V, A, D\}$. On the $x$-axis, we have the responses (Likert $1$--$9$), and on the $y$-axis, we have the frequency.}
% \end{figure*}

In this analysis, we employed Bayesian ordinal regression, using a cumulative model (for an introduction to Bayesian analysis, see~\citep{furia20bda}). One could imagine two other potential models, i.e., the sequential model or the adjacent category model. However, since Likert ($1$--$9$) scales were used for the outcome, cumulative models are more suitable, i.e., the sequential model would be suitable if we want to analyze the number of correct designs predicted from experience. In contrast, the adjacent category model would be appropriate if we want to predict the number of correctly solved sub-items of a complex task---none of this was of interest to us~\citep{burknerV19ordreg}.

Several models were designed, and their relative out-of-sample prediction capabilities were evaluated iteratively. The final model, below, includes all relevant predictors and has the same out-of-sample capabilities as other comparable models. For model comparison, we used state-of-the-art model evaluation~\citep{vehtariGG17loo}.\footnote{Pareto $k < 0.5$ and $\mr{LOOIC} = 2406.0$.} 

Next, follows the design of the final model and the corresponding priors. If we want to make a comparison with a frequentist approach, then one could claim that we have fixed and random effects in our model (i.e., a mixed-effects model); however, in a Bayesian setting, we use the term multilevel model, since that allows us also to employ hyperparameters with corresponding priors.

\begin{IEEEeqnarray}{rCl}
V_i,A_i,D_i & \sim & \mr{Cumulative}(\phi_i, \kappa) \\
\phi_i & \sim & \beta_1 \mr{EDUCATION}_i + \beta_2 \mr{EXAMPLE}_i + \beta_3 \mr{MAJOR}_i + \beta_4 \mr{ROLE}_i \\ 
& + & \beta_5 \mr{LANGUAGE}_i + \beta_6 \mr{ENTITIES}_i + \beta_7 \mr{EXPERIENCE}_i \\ 
& + & \beta_{\mr{SUBJECT}[i]} \\
\beta_1 & \sim & \mr{Dirichlet}(2,2,2,2,2) \\
\beta_{\mr{SUBJECT}} & \sim & \text{Half-Cauchy}(0,2) \\
\beta_2,\ldots, \beta_7 & \sim & \mr{Normal}(0,0.5) \\
\kappa & \sim & \mr{Normal}(0,5)
\end{IEEEeqnarray}

In the first line we model each outcome, $\{V, A, D\}$, using a cumulative likelihood. The parameters $\phi$ and $\kappa$ are the linear regression and the intercepts, respectively, which we model for each outcome (i.e., we have eight intercepts for each outcome since the outcome was Likert scale $1$--$9$).

In the next three lines, we have the linear regression. We have eight parameters we want to estimate, one for each of our predictors. The parameters $\beta_1$ and $\beta_{\mr{SUBJECT}[i]}$ are special as we will see next.

On Line $5$, we assign $\beta_1$ a Dirichlet prior. The Dirichlet prior is the multivariate generalization of the Beta distribution (a distribution commonly used to model a probability $[0,1]$). Using Dirichlet, we can model an array of probabilities; i.e., in this case, we model five probabilities and use a very weak prior (the $2$s), indicating that we do not have any prior knowledge. The reason we use a Dirichlet here is monotonicity, i.e., the predictor $\mr{EDUCATION}$ is an ordered categorical variable indicating the level of education. We, thus, want to model the probability separately for each of the categories in education.

Continuing on Line $6$ we assign $\beta_{\mr{SUBJECT}}$ a $\text{Half-Cauchy}(0,2)$ prior. This prior is common when modeling standard deviations and allows only positive real numbers ($\mathbb{R}^+$). To analyze variability in this way goes by many names, e.g., random effects or varying intercepts. The reason we use it is due to our following the latest recommendations by designing the experiment to collect within-person measurements~\citep{leek2017five}, i.e., each subject has been randomly allocated several tasks and, thus, we model the variability of each subject to partially pool the estimates, to avoid overfitting.

Proceeding to Line $7$, we assign the priors $\mr{Normal}(0,0.5)$ for the remaining parameters while, on the last line, we assign the prior $\mr{Normal}(0,5)$ to all intercepts for each outcome. (It is common to assign a broader prior for intercepts.)

The careful reader would react to what could be perceived as tight priors for several parameters, i.e., $\mr{Normal}(0,0.5)$. However, first, using $\mr{Normal}(0,0.5)$ on six parameters still makes an impressive standard deviation, $(6*0.5)^2 = 9$, and, second, the combination of all priors established \textit{a nearly uniform prior} on the probability scale, i.e., prior predictive checks and a sensitivity analysis were conducted. 

\begin{figure*}
\centering

\subfloat[Prior predictive check]{%
  \includegraphics[width=0.4\textwidth]{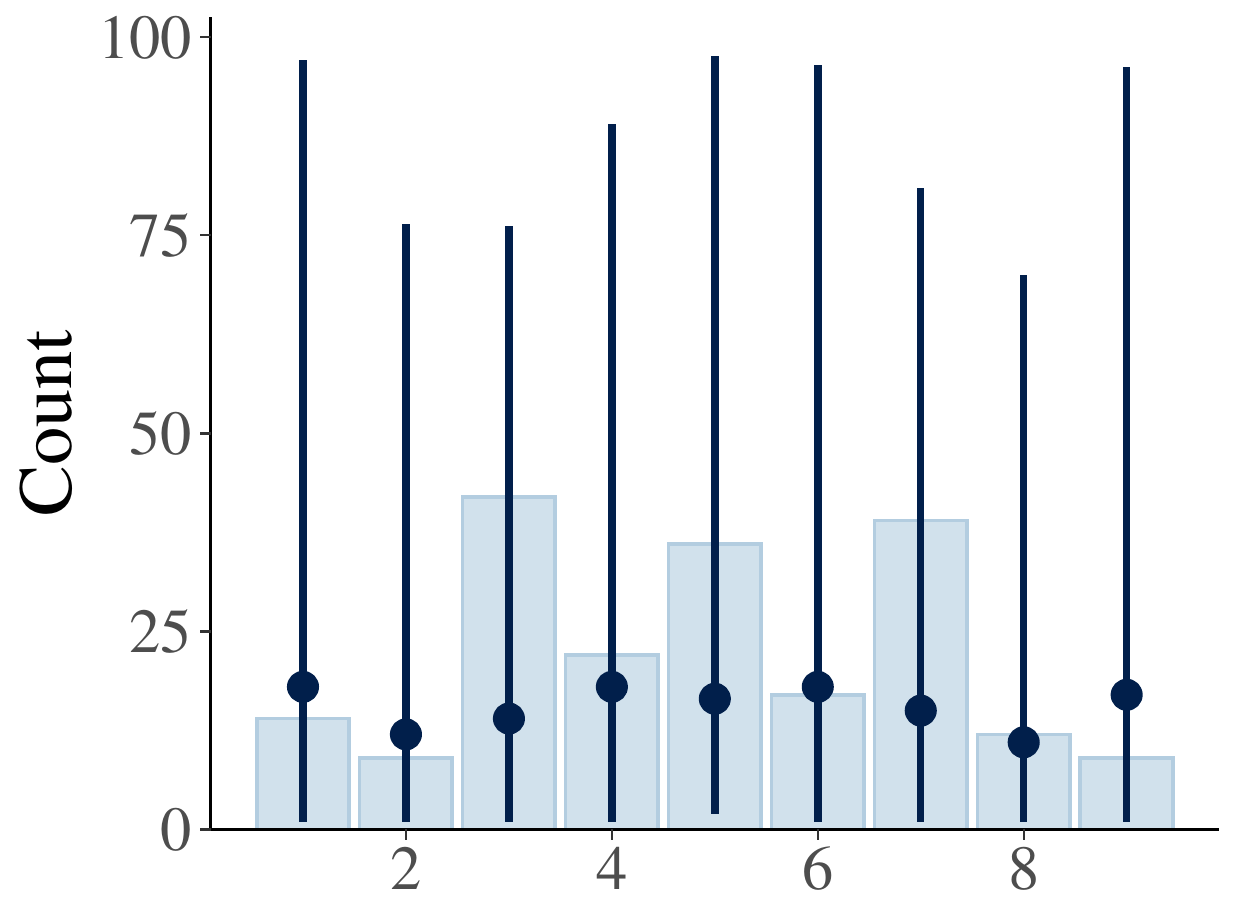}%
  \label{fig:pri}%
}\qquad
\subfloat[Posterior predictive check]{%
  \includegraphics[width=0.4\textwidth]{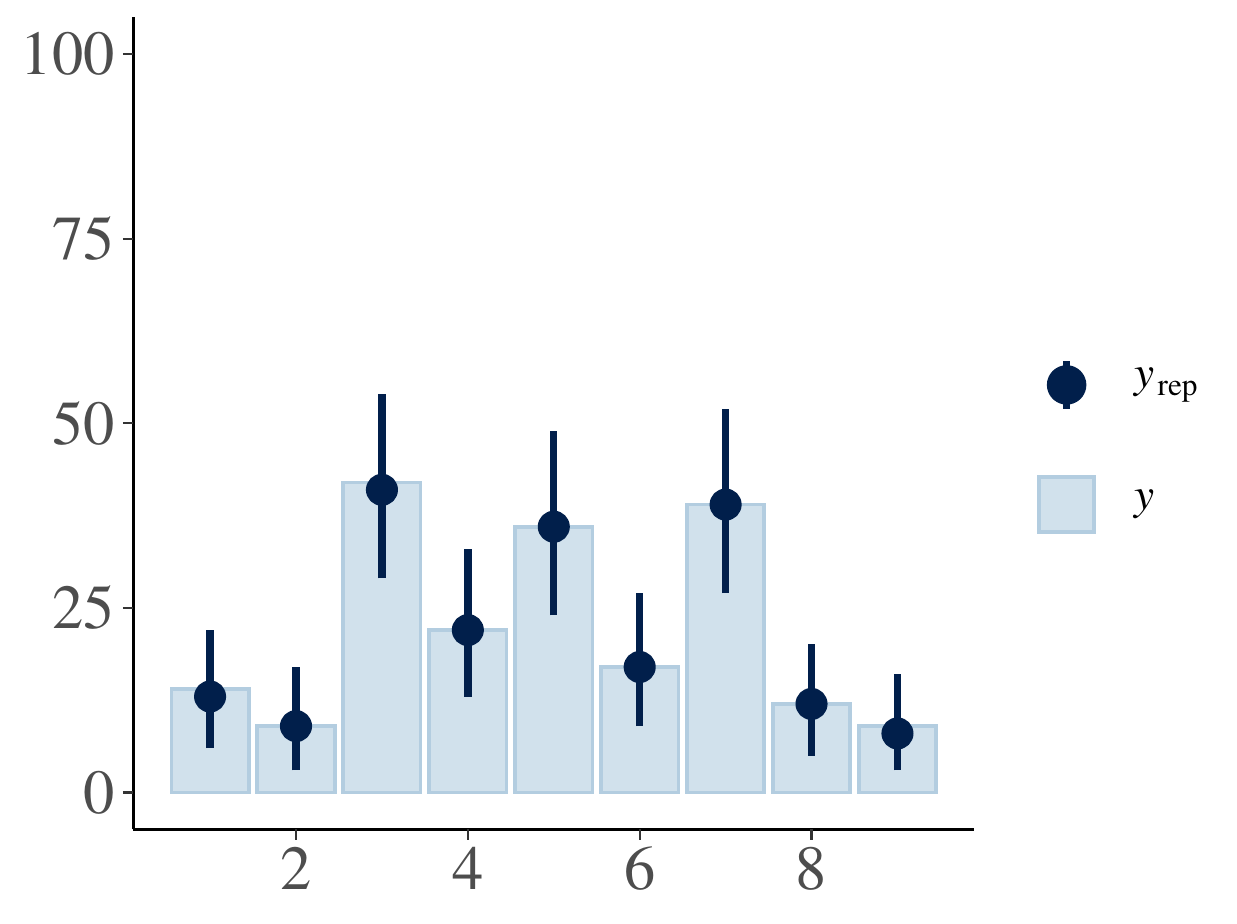}%
  \label{fig:post}%
}
\caption{Prior and posterior predictive checks ($y$ is the empirical data, and $y_{\rm{rep}}$ are $100$ draws from the prior (a) and posterior (b) probability distributions). The left plot shows the prior predictive checks (where no empirical data was used). The uncertainty is considerable (the lines), and the median values (the dots) are approximately the same for all items on the Likert scale, like it should be since only the priors are used. Compare this to the right plot, where we have drawn samples from the posterior probability distribution, i.e., we have fitted our model with data, the data has provided evidence, and thus the priors have been what is commonly referred to as `swamped', since the uncertainty has decreased.}
\end{figure*}

Since we used dynamic Hamiltonian Monte Carlo to sample, we also have several diagnostics. In our case, the model showed no indications of a biased posterior, and diagnostics ($\widehat{R}$, effective sample size, and trace plots) indicated that the chains had converged. Posterior predictive checks showed that the data swamped the priors (see \figsRef{fig:pri}{fig:post} for a visualization of the prior predictive checks and posterior predictive checks).

Continuing this section, we will next look at the output from the model. First, we will present the standard deviations for each outcome's random effects and any interesting population-level effects. Then, we will predict outcomes while fixating specific parameters. The final part will present the results of the hypothesis testing (Bayes factor).

Analyzing the variance, there is not much difference in the uncertainty of the estimates concerning $\sigma$ for our three outcomes, as the standard deviations' credible interval mass vary from $0.88$ ($\sigma_V$) to $1.1$ ($\sigma_A$). In short, the uncertainty for each outcome, $\{V,A,D\}$, is very much the same, but, notably, valence ($V$), has the lowest standard deviation $\sigma=0.39$, while arousal ($A$) has the largest standard deviation, $\sigma=0.87$, indicating more uncertainty in between-subjects variability. This can be interpreted as that the within-subject design and analysis we employed was beneficial (it was important to model different dispersions). 

% \begin{table*}
% \begin{center}
% \caption{Standard deviations of random effects. The standard deviation in $\sigma_A$ indicates more uncertainty compared to, e.g., $\sigma_V$, however it is alsoo clear that there is quite some uncertainty in the estimates themselves (the last two columns), indicating that considerable variance exists between subjects.}\label{tbl:sd}
% \begin{tabularx}{\textwidth}{XSSSS}
%     \hline
%      &  {Est.} & {Est.~Error} & {l-95\% CI} & {u-95\% CI} \\
%      \hline
%     $\sigma_V$ & 0.39 & 0.24 & 0.02 & 0.90 \\
%     $\sigma_A$ & 0.87 & 0.27 & 0.32 & 1.42 \\
%     $\sigma_D$ & 0.60 & 0.26 & 0.07 & 1.11 \\
%     \hline
% \end{tabularx}
% \end{center}
% \end{table*}

Analyzing the estimates, and their corresponding 95\% credible intervals, led to $5$ estimates being singled out as interesting (\tabRef{tbl:params}). Four were significant on the arbitrary 95\%-level (i.e., not crossing zero), while one is strongly positive, albeit not significant on the 95\%-level.

\begin{table*}
\begin{center}
\caption{Parameters of interest.}\label{tbl:params}
\begin{tabularx}{\textwidth}{XXSSSS}
    \hline
    Outcome & Parameter &  {Est.} & {Est.~Error} & {l-95\% CI} & {u-95\% CI} \\
     \hline
    Dominance ($D$) & EXAMPLE (BL) & -0.78 & 0.34 & -1.43 & -0.12 \\
    Valence ($V$) & EXAMPLE (BH) & 0.73 & 0.34 & 0.07 & 1.39 \\
    Valence ($V$) & EXAMPLE (DL) & -0.83 & 0.35 & -1.52 & -0.14 \\
    Valence ($V$) & EXAMPLE (CL) & 0.72 & 0.36 & 0.02 & 1.42 \\
    Valence ($V$) & EXPERIENCE & 0.25 & 0.16 & -0.05 & 0.56 \\
    \hline
\end{tabularx}
\end{center}
\end{table*}

Since Experience has much probability mass on one side of zero ($[-0.05;0.56]$), we will analyze it further to understand its predictive ability better. Before we analyze Experience further, let us look at the role Entities (i.e., the complexity of each task) has on the outcome. If it is not positive, then one could argue that they have had the wrong effect.

% If we draw $100$ samples from our posterior probability distribution, we will receive a better feeling for the uncertainty concerning the four significant parameters, while fixating all other parameters. In Fig.~\ref{fig:paramdraws}, we have set our parameters to median values (or the reference category in our sample), i.e.,

% \begin{figure}
% \centering

% \subfloat[]{\includegraphics[height=2.2in, width=0.3\textwidth]{figure/plots/outcome-V-sim.pdf}}\quad%
% \subfloat[]{\includegraphics[height=2.2in, width=0.3\textwidth]{figure/plots/outcome-A-sim.pdf}}\quad%
% \subfloat[]{\includegraphics[height=2.2in, width=0.3\textwidth]{figure/plots/outcome-D-sim.pdf}}
% \caption{From left to right $100$ samples drawn from our posterior probability distribution for the outcomes Valence (a), Arousal (b), and Dominance (c). On the $x$-axis, we move from Low to High, while the $y$-axis displays the Likert scale response. This way, we simulate the outcomes given a set of fixed values. Even though these parameters were significant, if fixated to a representative sub-sample it shows that there is much uncertainty, albeit we see some positive trends.}\label{fig:paramdraws}
% \end{figure}

% \begin{itemize}
%     \item Major: Software engineering ($n=90$)
%     \item Educational level: Bachelor ($n=85$)
%     \item Experience: $8$ (median)
%     \item Role: Developer ($n=125$)
%     \item Language: C\verb|#| ($n=100$)
%     \item Number of entities (complexity): $5$ (median)
% \end{itemize}

\begin{figure*}
    \centering
    \includegraphics[scale=0.25]{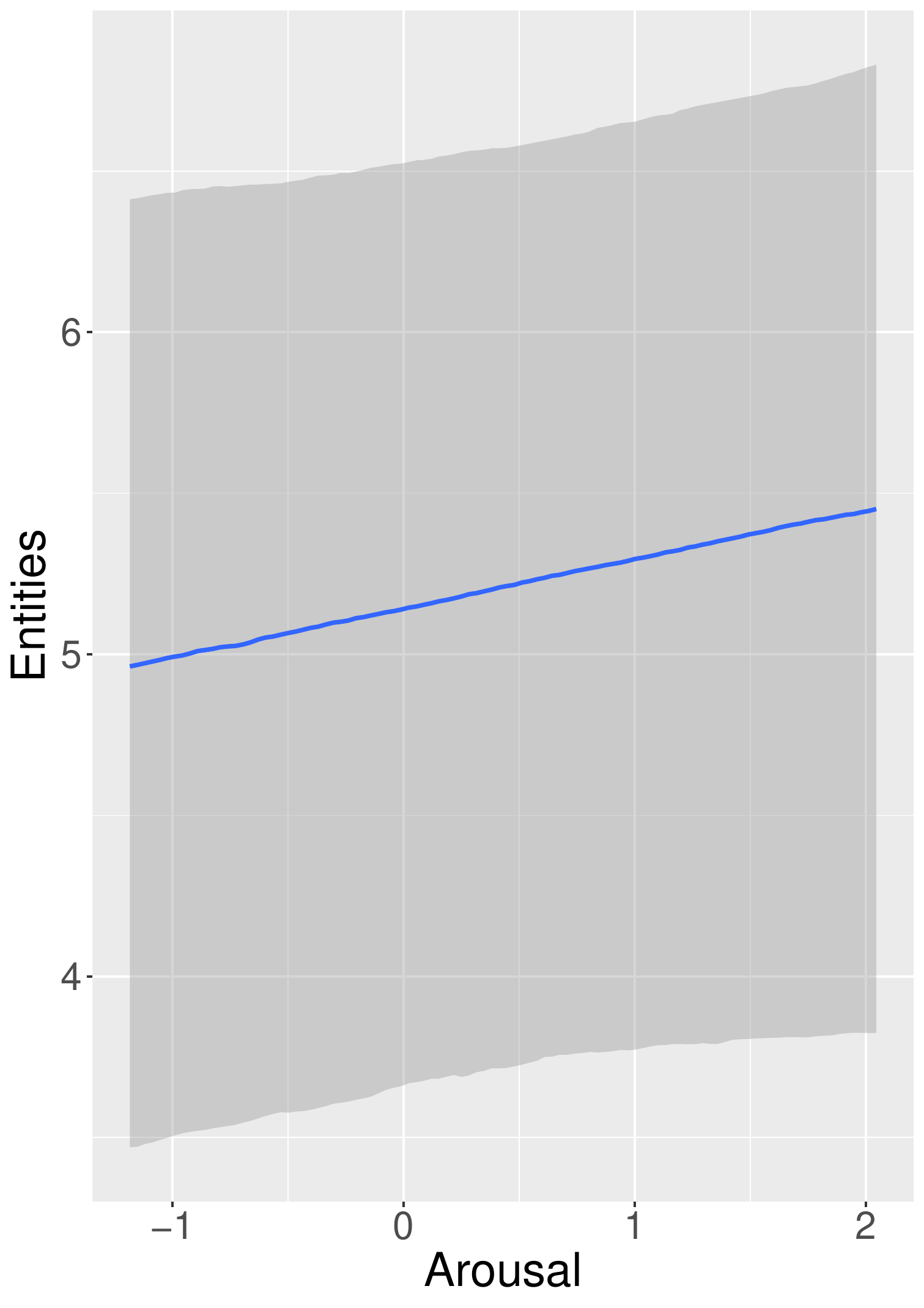}
    \caption{Conditional effect of Entities in the model. The more complex an entity (i.e., the more to the right we move on the $x$-axis), the higher the outcome on the Likert scale ($y$-axis). In this case, we looked at the outcome $A$ (arousal), but the same trend is visible in all three outcomes. The $x$-axis has been scaled, with $0$ corresponding to median complexity. (The line is the median outcome, while the gray area is the 95\% uncertainty around the median.)}
    \label{fig:ent}
\end{figure*}

To investigate Entities we need to determine what covariate values to use. One possible way to do this is to set all values to their mean for continuous variables, while the reference category is used for factors, and then examine the conditional probabilities our posterior probability distribution provides us with. In \figRef{fig:ent}, we see a positive trend, which indicates that the model has been able to capture the role that complexity plays correctly.

\begin{figure*}
\centering

\subfloat[Valence]{%
  \includegraphics[scale=0.25]{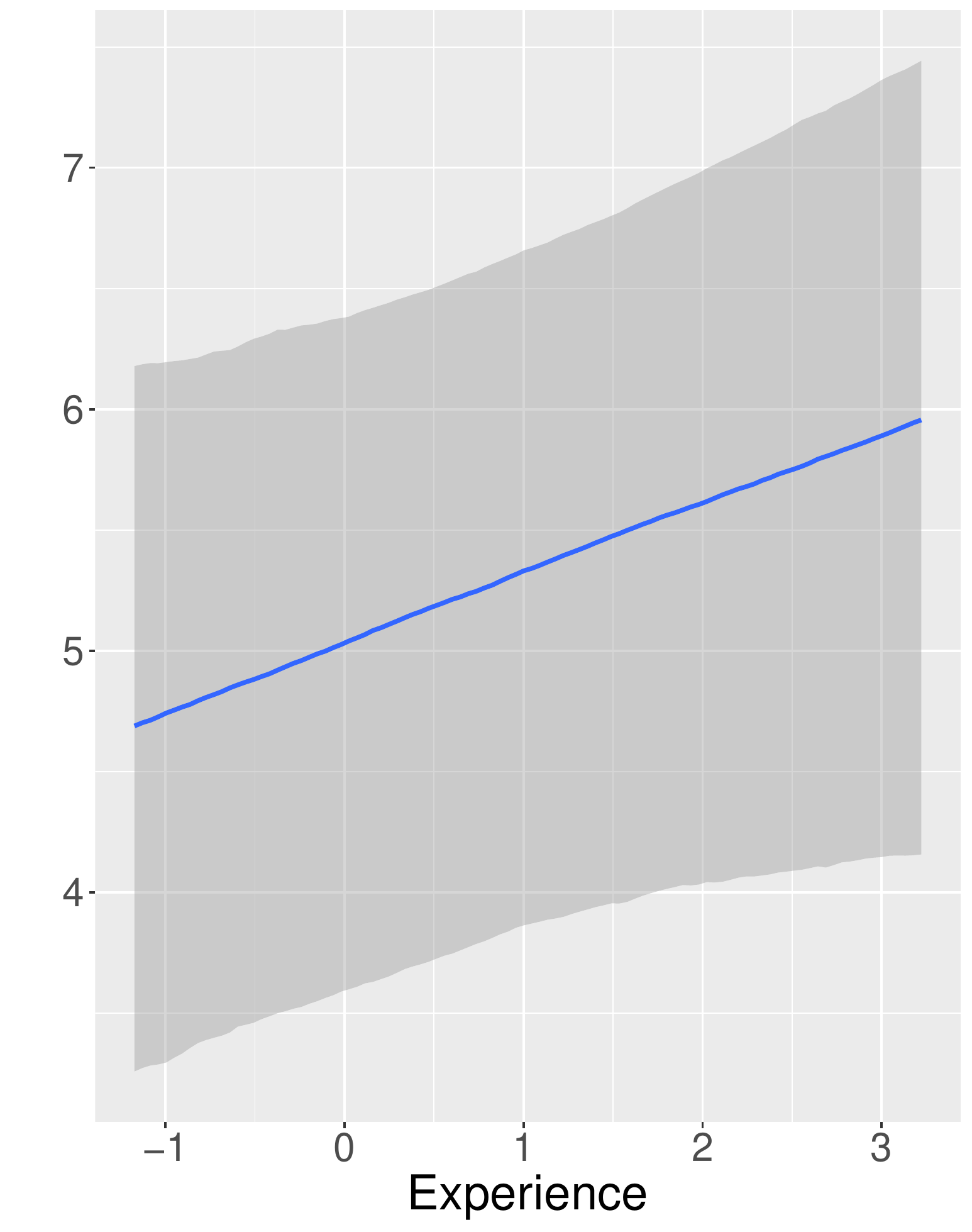}%
  \label{fig:v-exp}%
}\qquad
\subfloat[Arousal]{%
  \includegraphics[scale=0.25]{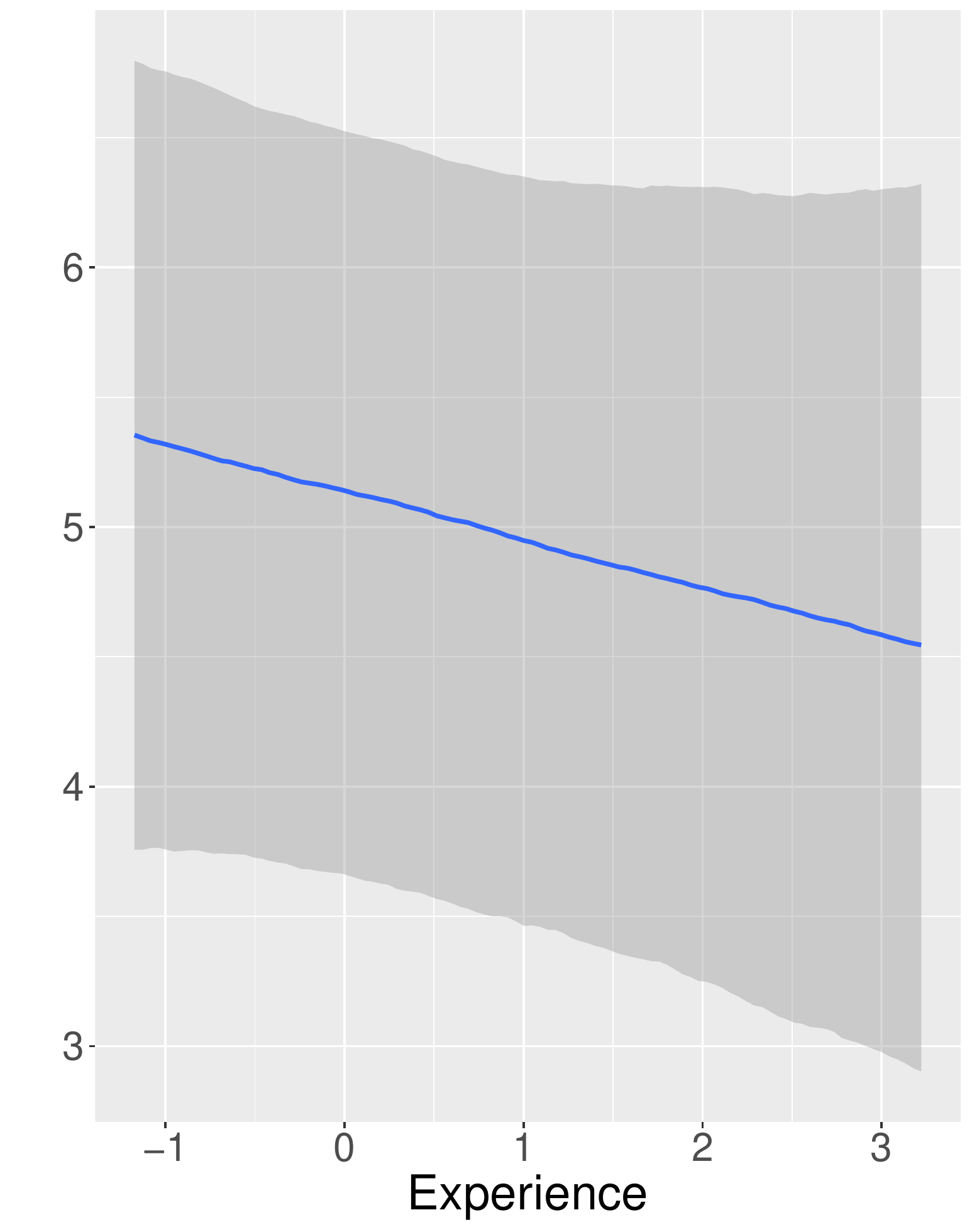}%
  \label{fig:a-exp}%
}\qquad
\subfloat[Dominance]{%
  \includegraphics[scale=0.25]{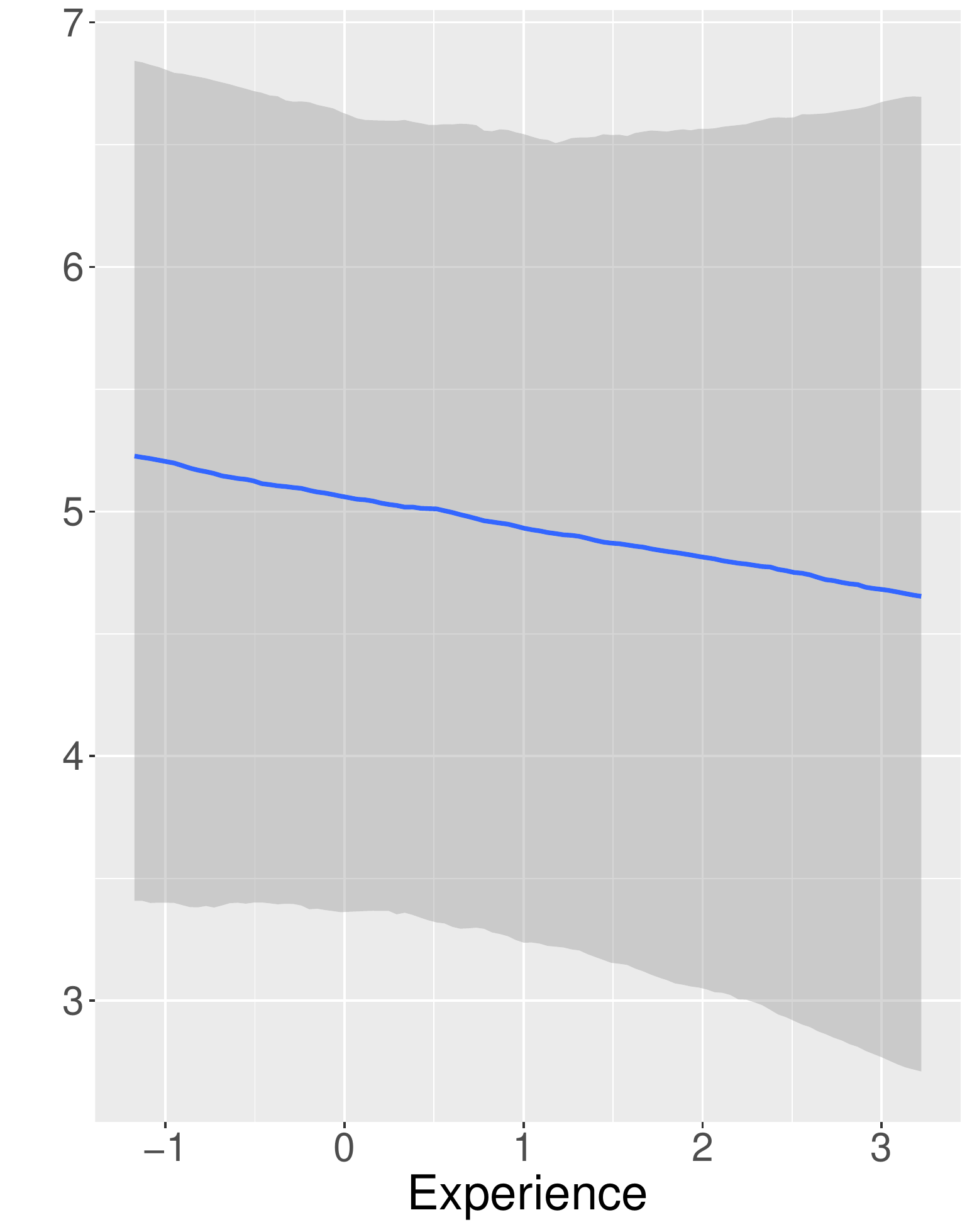}%
  \label{fig:d-exp}%
}
\caption{An overview of the conditional effects on Experience, given our three outcomes $\{V,A,D\}$. Lines correspond to the median, while the gray area is the $95$\% credible interval. For valence ($V$), we have a positive effect, i.e., the more experienced the subject, the higher the response on the Likert scale, while the opposite holds for Arousal ($A$) and Dominance ($D$).}
\end{figure*}

Finally, we would like to see the role Experience plays by analyzing it more carefully. If we turn our attention to \figsRef{fig:v-exp}{fig:d-exp}, we see that the role it plays differs, depending on our outcome. For Valence ($V$), we have a positive effect, i.e., the more experienced the subject, the higher the response on the Likert scale, while the opposite holds for Arousal ($A$) and Dominance ($D$). Here, it is crucial to keep in mind the direction of the SAM, i.e., an increase in $V$ score means more displeasure; arousal increases as $A$ decreases; low $D$ scores denote submissiveness.

Having analyzed the conditional effects, we now turn our attention to measuring the strength of the evidence we have gathered. Our tests will \textit{not} examine the significant population-level effects, which we list in \tabRef{tbl:params}; after all, we know that they are significant on the traditional $95$\%-level. Instead, we will focus on the contrasts between Low ($L$) and High ($H$) settings for our predictor Example. This means that we can present the results as several hypothesis tests ($5$ artifacts times $3$ outcomes equals $15$ tests in total). Since we have a posterior probability distribution, we do not have to, generally speaking, worry about multiple tests, which is often the case in a frequentist setting~\citep{gelmanT00typeS,gelmanHY12worry}.

\begin{table*}
\centering
\caption{Decision thresholds for hypothesis testing using Bayes factor, according to~\citet{Kruschke2010}.}\label{tbl:bf}
\begin{tabularx}{\textwidth}{XXl}
\hline
Symbol & Evidence ratio & Description\\
\hline
** & $>10$ & Strong evidence for H1 \\
* & $3$--$10$ & Moderate evidence for H1 \\
? & $1$--$3$ & Anecdotal evidence for H1 \\
? & $1/3$--$1$ & Anecdotal evidence for H0 \\
* & $1/30$--$1/10$ & Moderate evidence for H0 \\
** & $<1/10$ & Strong evidence for H0 \\
\hline    
\end{tabularx}
\end{table*}

For hypothesis testing, we will use Bayes factor to avoid the usage of $p$-values and, thus, to receive verdicts both in favor of and against a given hypothesis~\citep{goodman99p,goodman99bayes}. For our accept\slash reject decisions, we follow recommended practices as presented in \tabRef{tbl:bf}~\citep{Kruschke2010}.

Our hypothesis tests were unidirectional and, thus, tested that Low $<$ High, e.g., 
\begin{equation*}
H_0: \mr{Example}_{\mr{AL}} < \mr{Example}_{\mr{AH}},
\end{equation*}
\noindent which is to be interpreted as Example $A$ Low is less than Example $A$ High (and we analyze this inequality for each of our outcomes $\{V,A,D\}$).

\begin{figure*}
\centering

\subfloat[Valence ($V$)]{%
  \includegraphics[width=0.5\textwidth]{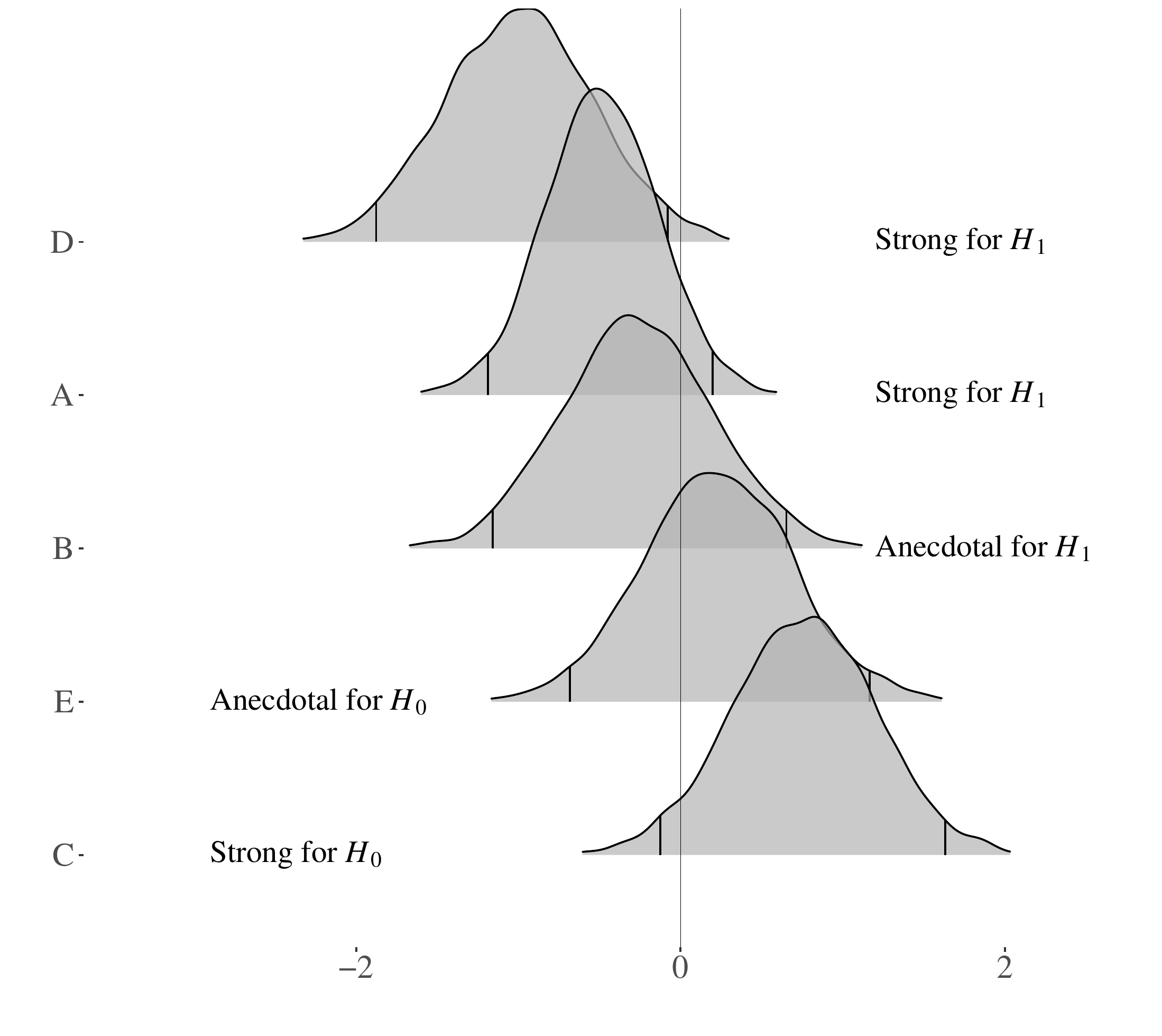}%
  \label{fig:hypo-v}%
}
\subfloat[Arousal ($A$)]{%
  \includegraphics[width=0.5\textwidth]{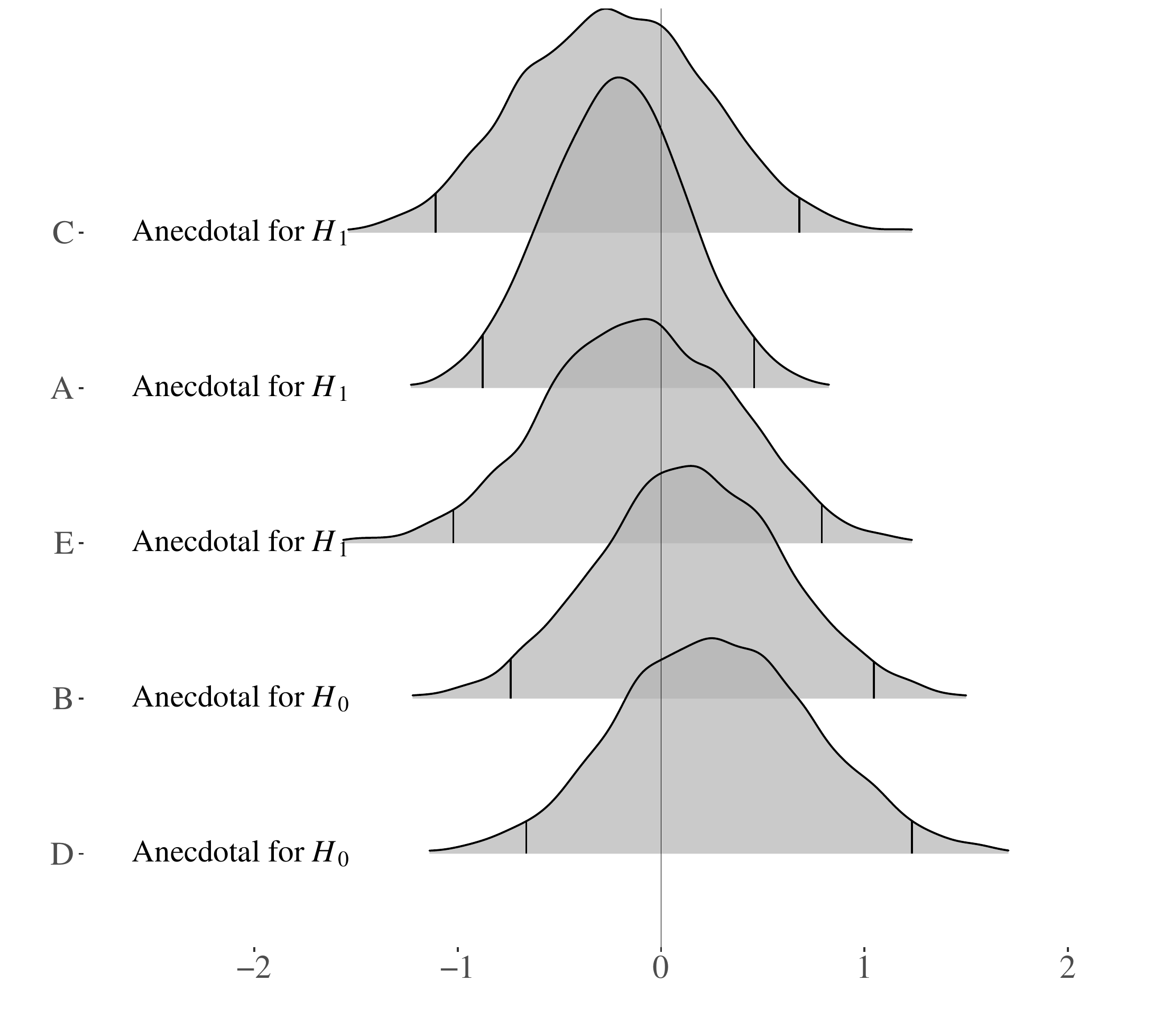}%
  \label{fig:hypo-a}%
}\\
\subfloat[Dominance ($D$)]{%
  \includegraphics[width=0.5\textwidth]{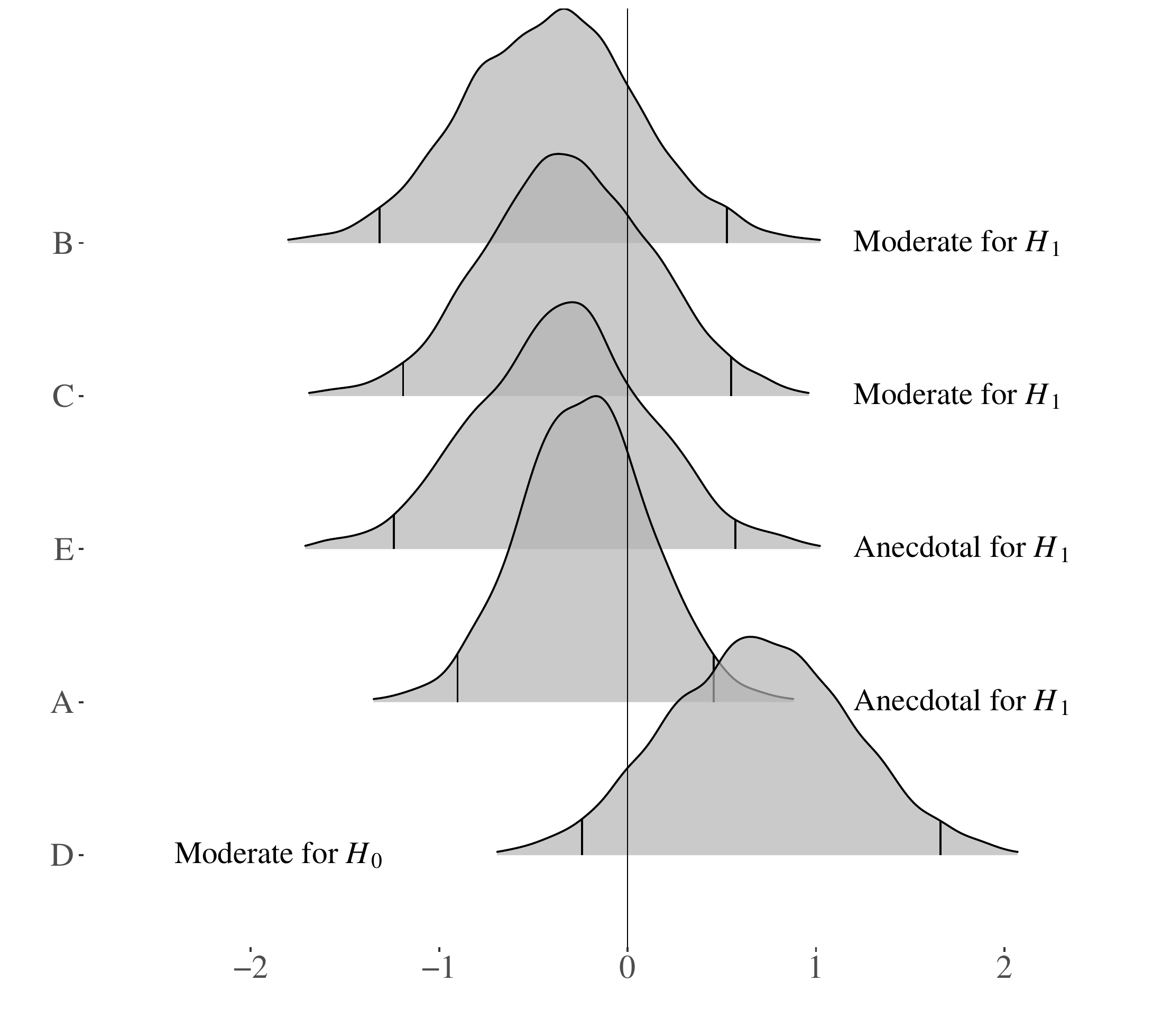}%
  \label{fig:hypo-d}%
}
\caption{A visual overview of all hypothesis tests, given our three outcomes $\{V,A,D\}$ ($x$-axis is the contrast). On the $y$-axis, Examples ($A$--$D$) are ordered according to the direction of evidence \textit{starting with the most negative direction}. Next to each distribution, a short note clarifies the results of the tests (according to \tabRef{tbl:bf}). Finally, the distributions have 2.5\% and 97.5\% quantiles drawn in the tails. As an example, artifacts $D$, $A$, and $C$, in outcome $V$ (valence) indicate strong evidence. In the two former cases we have \textit{strong} evidence \textit{for} $H_1$, while in the latter case we have strong evidence \textit{for} $H_0$.}
\end{figure*}

If we plot the posterior probability distributions for each hypothesis test ($15$ in total), one can perhaps better see what a `significant' effect means in the context (\figsRef{fig:hypo-v}{fig:hypo-d}).

\subsection{Effect Sizes}
Looking at \figsRef{fig:hypo-v}{fig:hypo-d} one sees three hypotheses that indicate strong evidence, i.e., Examples $D$, $A$, $C$ in outcome $V$ (valence). In the two former cases, we have evidence for $H_1$, while in the latter case we have evidence for $H_0$. Analyzing the effect sizes for these results is wanted. However, we also see two more results that could potentially also be of interest.

In \figRef{fig:hypo-d}, one can see that there are some probability distributions classified as providing \textit{moderate} evidence for $H_1$ or $H_0$, respectively (but they are still fairly close to a quantile). These are Examples $B$, $C$, and $D$. Even though we do not have strong evidence speaking in favor (or not) of a hypothesis, it could be of interest to see what this entails concerning effect size.

In short, we want to see, on average, how large an effect size it would be to move from $H$ to $L$ for each of the six Examples. By drawing samples from our posterior probability distribution, we can easily compare the difference between levels. We leave all variables according to what we have in the sample (e.g., the distribution concerning Experience is the same) and vary only the Example level to see what this means on the outcome scale. \tabRef{tbl:es} provides us with an overview of the six effect sizes.

\begin{table*}
    \centering
    \begin{tabularx}{\textwidth}{Xlrrrrr}
         \hline
         Outcome & Example &  {Min.} & {1st quant.} & {Median} & {3rd quant.} & {Max.}\\
         \hline
         %-3.550  -1.500  -1.000    -0.550   1.750
         Valence ($V$) & $D$ & $-3.6$ &   $-1.5$  & $-1.0$ &  $1.8$  & $3.9$\\
         %-2.9500 -1.0000 -0.5000   0.0000  2.0500
         Valence ($V$) & $A$ & $-3.0$ & $-1.0$ & $-0.5$ & $0.0$ & $2.0$\\
         %-1.6667  0.2222  0.7500   1.2778  3.7778
         Valence ($V$) & $C$ & $-1.6$ & $0.2$ & $0.8$ & $1.3$ & $3.8$\\
         \hline
         %-3.2105 -0.9474 -0.4211   0.1053  2.6842 
         Dominance ($D$) & $B$ & $-3.2$ & $-1.0$ & $-0.4$ & $0.1$ &  $2.7$\\
         %-3.2778 -0.8889 -0.3333   0.2222  2.6111
         Dominance ($D$) & $C$ & $-3.1$ & $-0.9$ & $-0.3$ & $0.2$ &  $2.6$\\
         %-2.5000  0.2000  0.7750   1.3000  3.8000
         Dominance ($D$) & $D$ & $-2.5$ & $0.2$ & $0.78$ & $1.3$ &  $3.8$ \\
         \hline
    \end{tabularx}
    \caption{Raw effect sizes from posterior samples ($10,000$ draws) of the posterior predictive distribution. These samples have higher variance than samples of the means of the posterior predictive distribution since residual error is incorporated.
    The first three rows present raw effect sizes where the hypothesis test found strong evidence, while the last three rows show where there was moderate evidence. The median column is the size of the effect (on the outcome scale) for the contrasts $L$--$H$. If we look at the first row we see an effect size of $-1.0$, i.e., the difference between Low--High, for Outcome $V$ and Example $D$, is $-1.0$ on the Likert scale with the quantiles $[-1.5,1.8]$. This should not be confused with the hypothesis tests we conducted (~\figsRef{fig:hypo-v}{fig:hypo-d}), which tested if Low$<$High.}
    \label{tbl:es}
\end{table*}

One can conclude this section by claiming that we have some interesting effects, some even based on substantial evidence. These are summarized in the box below as findings F1--F11.

\clearpage

\RQBox{
    \textbf{Findings for RQ1:} 
    \begin{itemize}
        \item[F1] Cyclically-dependent modularization (ScD-H) is less pleasant than its refactored (ScD-L) counterpart (strong evidence).
        \item[F2] Missing encapsulation (ScA-H) is less pleasant than its refactored (ScA-L) counterpart (strong evidence).
        \item[F3] Broken modularization (ScC-H) is \emph{more} pleasant than its refactored (ScC-L) counterpart (strong evidence).
        \item[F4] Missing Hierarchy (ScB-H) is, likely, \emph{less} dominating than its refactored (ScB-L) counterpart (moderate evidence).
        \item[F5] Broken modularization (ScC-H) is, likely, \emph{less} dominating than its refactored (ScC-L) counterpart (moderate evidence).
        \item[F6] Cyclically-dependent modularization (ScD-H) is, likely, more dominating than its refactored (ScD-L) counterpart (moderate evidence).
    \end{itemize}
    
    \bigskip
    
    \textbf{Findings for RQ2:} 
    
    % The conditional effect of work experience (once setting all values to the median or reference category) is:
    
    % \begin{itemize}
    %    \item Valence: Positive (i.e., the more work experience, the higher the Likert value).
    %    \item Arousal: Negative (i.e., the more work experience, the lower the Likert value).
    %    \item Dominance: Negative.
    %\end{itemize}
    
    % Of the three effects above, Dominance provides moderate evidence, while the other affects can be seen as anecdotal. The effects are further plotted in Figs.~\ref{fig:v-exp}--\ref{fig:d-exp}.
    
    \begin{itemize}
        \item[F7] Work experience, likely, correlates with submissiveness (moderate evidence).
    \end{itemize}
        
    \bigskip
    
    \textbf{Additional findings:}
    \begin{itemize}
    \item [F8] Refactored Missing Hierarchy (ScB-L) yielded particularly submissive responses.
    \item [F9] Missing Hierarchy (ScB-H) yielded particularly displeasing responses.
    \item [F10] Refactored Cyclically-Dependent Modularization (ScD-L) yielded particularly pleasing responses.
    \item [F11] Refactored Broken Modularization (ScC-L) yielded particularly displeasing responses.
    \end{itemize}
    
    % The above effects are further elaborated on in Figs.~\ref{fig:hypo-v}--\ref{fig:hypo-d}.
}

\section{Qualitative Analysis and Results \label{sec:qualitative}}
Analyzing the data set (which predominantly concerned the participants' general experience of TD, rather than the experiment scenarios) revealed that the participants have strong and negative affects toward TD and are inclined to talk about their reactions. Their argumentation was clearly of the stimulus-response variety, i.e., they viewed TD as an action they are exposed to, leading to counteractions. The participants' discussions centered around what one might think of as defense or coping mechanisms for said stimulus.\footnote{These are established terms within psychology, and the surrounding theory could not be delved into for the scope of this study. In this article, we will instead use the term \psychologicalRebound{} to avoid overloading the terms.}

\begin{figure}
    \centering
    \includegraphics[width=\linewidth]{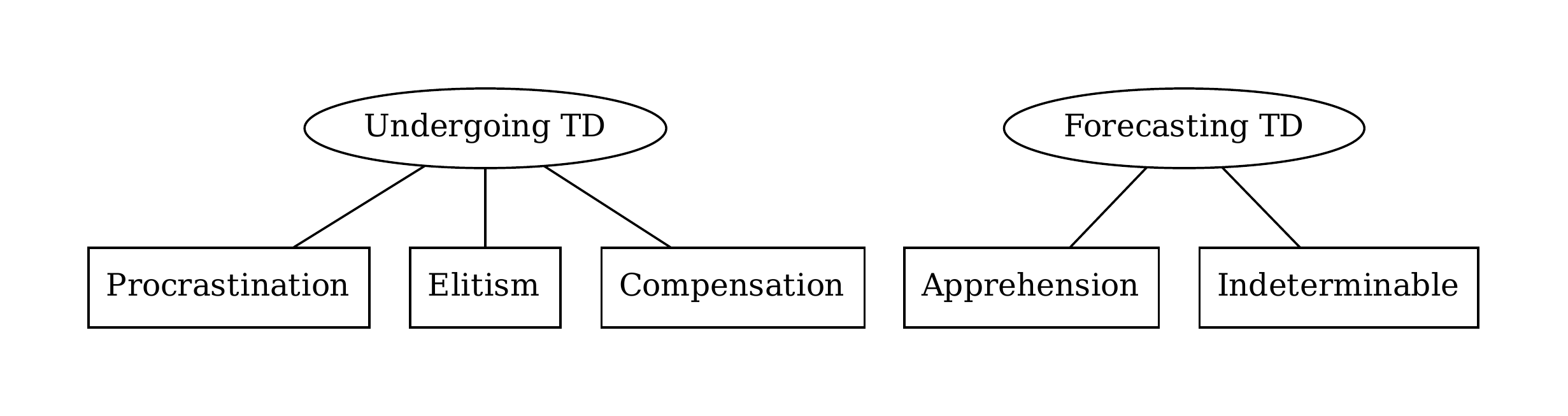}
    \caption{Thematic map of how software practitioners reason about TD in tandem with affects.}
    \label{fig:thematicMap}
\end{figure}

The thematic map (including two themes and five sub-themes) constructed during the analysis is included in \figRef{fig:thematicMap}. The first theme (three sub-themes) describes the participants' reflections (with regard to affective state) on undergoing TD intense areas (\undergoingTD{}), e.g., encountering TD, when working with some other task.

Among its sub-themes, we first consider \procrastination{}. At its core, this sub-theme is about instances where practitioners try to delay or avoid dealing with the debt or its consequences. Often, this is related to the sense of feeling overwhelmed when facing TD\@. 

\procrastination{} may surface in several different forms. For example, one interviewee reported that TD could cause task abandonment. \interviewQuote{the more, like, bad code I see [in the same place], the more, like, bored and [indifference] \quoteSkip{} It's like, \innerQuote{[vocable of quitting], I give up}. It's like, \innerQuote{it's too much now, I give up.}}

This feeling of resignation was echoed by another practitioner, who also suggests that tightly coupled code is cognitively taxing. \interviewQuote{it had this \texttt{\emph{instanceof}} bit that implies that it knows about something else, so then you have to start knowing about two places at once, in parallel, and that usually gets super messy. [vocable of distaste] Yeah, so it's, sort of, being in control and being able to fix it.}

At the same time, \procrastination{} is not constrained to low levels of arousal. Quite the opposite, in some instances, it can lead to an impulsive and risky overhaul of parts of the codebase: \interviewQuote{I would throw away and rewrite it}.

From these examples, it is clear that TD can cause \psychologicalRebound{} effects that are harmful to the software project in ways that go far beyond the human aspects perspective. For example, abandoning tasks because of TD can upset backlog prioritization or result in project slippage. Similarly, the urge to overhaul the code base could, e.g., invalidate prior trade-off analyses.

Unsurprisingly, the participants were aware of the consequences and severity of \procrastination{}. One interviewee said, \interviewQuote{I think the detrimental part is when you feel like you don't wanna touch it \quoteSkip{} even if I do touch it in the end, it will take a longer time before I actually dare.}

Next, the second sub-theme is \elitism{}. It encompasses reactions to TD, violating some expectations that one holds oneself, one's colleagues, or the code base to. In the case of \elitism{}, these expectations typically do not represent a shared set of values and beliefs among the parties. Hence, the discourse in this sub-theme was notably flavored by negative interpersonal dynamics.

\elitism{} is reflected in several different affects that appear to fall on a wide scale of blaming the author of the code. One example of a low amount of blame was one participant who expressed disappointment. \interviewQuote{
if you have a great design, a great architecture, following the SOLID principles. That are loosely coupled. [Then,] they [code problems] are easy to fix. The problems. Easy to change. That is the most important, to me. So there are some---they are fundamentals of how I think when I design a program. So [code] violating those principles make me feel very sad.}

As can be seen, this suggests that the code base itself influenced the participant's affects, i.e., more or less decoupled from its author. On the other hand, another interviewee, who experienced distrust, accentuates the author's (perceived) skill and does not separate it from the quality of the code: \interviewQuote{I've seen things where people mix really bad indentation, combined with not having, like, opening and closing brackets for \texttt{\emph{for}}-loops, for example. Using, like, short notation. We can have, like, one-liners after \texttt{\emph{if}}-statements, for example. I mean, those things are just terrible, 'cause you don't know what belongs where. It's messy and there are, like, no, like, blank lines between---additional spacing between things or anything. It's just a bunch of code, with wrong indentation. Sometimes indented, sometimes not. And unclear what belongs to which statements. \quoteSkip{} it's easier to spot it [than architecture]. And it's so, like, something I really think people should know how to do. It's so basic, in programming. So, yeah, I think so. It makes me a bit more worried, so to say, when I see that stuff. 'Cause it's very much easier to do correctly.}

Continuing on this blame scale, examples arise where the code is de-empathized in favor of focusing on its author. For instance, one interviewee expressed scorn and a notion of coding style reflecting one's personality. \interviewQuote{I get a bit annoyed with people that try to be too smart with the programming language. They know, like, a short way of writing things, and they know exactly what happens. \quoteSkip{} So I'm more for, like, writing simple, easy to understand code. So that everyone that follows, can easily make changes to it. So yeah, that annoys me a bit: when people try to be too clever. They wanna show off that they are smart, by using, like, weird functions of a language.}

Viewed together, these examples suggest that \elitism{} may arise from the misalignment of quality expectations. However, this is perhaps not obvious to the practitioners, as the focus is not on addressing these alignment problems. Clearly, \elitism{} threatens to cause conflict among employees, but can also rationalize TD by acknowledging the debt as a result of business constraints: \interviewQuote{However, I also feel that when I read someone else's code, that's really bad---or shit, or something---I also realize that this might have been done under pressure, depending on the project and stuff. So I accept these technical debts better. Unless it's just plain bad and not time-saving at all.}

\elitism{} can be dangerous also when it does not result in (external) conflict. One interviewee highlights the risk of it causing high levels of stress. \interviewQuote{Yeah, this was people that were sort of in the more, like, architect roles, usually. Then they put on too much work on their shoulders. They were the guys that always wanted to do everything by themselves. And, sort of, tended to burn out after a while, 'cause they just had too much to do. You could see that they were stressed about it [soft deadlines].}

The final sub-theme of the first theme is \compensation{}\footnote{In the behavioral (not the financial) sense of the word.}, which concerns constructively addressing TD\@. Often, the TD items are viewed as opportunities for improvement. As one interviewee put it, \interviewQuote{So, there definitely is this scope for improvement, but I would not call anybody else's code as poor. \quoteSkip{} I generally do not get any negative feelings about it [code clones]. But I do look at it as an opportunity to improve the code myself.}

\compensation{} is not limited to correcting an instance of TD, but it can encourage preventive actions, e.g., informing the code author about their mistake: \interviewQuote{Personally, I would use \texttt{\emph{git blame}} to see who wrote the code and then, if I can contact them, I say \innerQuote{okay, next time, you should do it better. Because this, like, it may take a lot of time for others to trace their issues.}}

One interviewee even suggested that affects can be leveraged to improve the code base, as they can act as software quality proxies: \interviewQuote{emotions aren't bad or good. If a team member is that mad about something, I just use that as an indicator that something is bad in the code. So that person is right to be angry, and we can use that to either fix it, or use that as an argument for---like, in the future---like, let's refactor this in the next sprint, or whatever.}

Together, these extracts show that \compensation{} is related to TD management and, more specifically, tactics for addressing TD maturely or constructively. Please note how these tactics are concerned with the practitioners' dominance concerning the code base. As one participant said, \interviewQuote{I want to rewrite it [code with inheritance issues]. \quoteSkip{}  to improve it and to just, yeah, maybe so I don't feel stressed about it. So I have control.}

So far, we have presented the components of the first theme. Before continuing with the next theme, the interactions between these components should be analyzed. Note how all three sub-themes appear to be \psychologicalRebounds{} for TD, albeit as different manifestations. \procrastination{} looks like an impulsive and naive, almost childish reaction to TD, where the practitioner does not acknowledge the consequences of their actions. These traits can also be seen in \elitism{}, but with regard to collaboration and teamwork rather than how the debt itself is approached. On the other hand, \compensation{} appears to be a manifestation of thoughtful consideration of how to manage the TD\@.

The second theme (two sub-themes) describes the participants' reasoning (with regard to affective state) when forecasting the consequences of TD (\forecastingTD{}), e.g., the effects of leaving TD unaddressed.

Its first sub-theme is \apprehension{}, which includes the anxiety of expecting future maintainability issues. A significant part of this sub-theme constitutes the participants' concerns about the extra psychological toll caused by TD\@. This toll can emerge when the practitioner believes there is a risk of the code leading to system failure. As one participant said, \interviewQuote{The spontaneous feeling was a bit stress about too much stuff going on. Too many components, and some strange dependencies. And too much inheritance. \quoteSkip{} Why [do I feel stressed]? 'Cause I can see myself maintaining that code. And I can see that code breaking in the long term. \quoteSkip{} 'Cause I don't want the system to break.}

This kind of uncertainty was echoed by another interviewee, who emphasized the toll of unforeseen consequences (ripple effects): \interviewQuote{for me, it comes back to, like, the control. I know that if I'm gonna touch this, I'm gonna pull a string, and then there's gonna come, like, a spider web with a spider in it. \quoteSkip{} You know that when you do something here, it's gonna affect something else.}

However, \apprehension{} is not limited to the technical considerations, as the psychological toll can also appear in the presence of tight schedules. As one practitioner put it, \interviewQuote{If you have time pressure to do something, and then you also know that you're in---I mean, \simpleQuote{this is gonna be hard to test. And to deliver it in time is gonna be tough.} Then it's super stressful. But if you don't have that pressure again, then it's easier again.}

Clearly, \apprehension{} is found in situations where the practitioner's dominance is on the submissive part of the scale, where they have low confidence in the code. Further, the extracts suggest that work tasks and business considerations are difficult to separate from their affective states. As one participant said, \interviewQuote{I mean, they [the technical and emotional viewpoints] are connected somehow. But through my years---my experience---I see a lot of problems with code violating these [SOLID design] principles. And that causes frustration when you try to fix bugs, improving the code, extend the code. So, it's more from a technical perspective, but they cause negative emotions.}

The last sub-theme is \indeterminable{}, which encompasses the difficulty of decoding TD\@. That is, understanding or sharing one's understanding of the TD in the system appears to be a non-trivial matter, which could play a key role in assigning value to TD items.

In industry, TD items are sometimes so opaque that professionals may not recognize them until they have paid a significant amount of interest. As one interviewee said, \interviewQuote{one time I was just gonna write some test for a thing we did. Then I realized the whole thing was such a debt-cluster that I just had to throw it away. I spent like three, four hours trying to help my team out. I didn't realize I did zero value [laughs] with that time.}

At the same time, TD might be widespread in the system, becoming a sort of background noise challenging to pinpoint. As one participant put it, \interviewQuote{sometimes you actually encounter some area that makes you really unhappy to be in. But then you also have these overarching stuff, that isn't really bothering you that much. But you always know it. You know it's always there. So it's way---it's less tangible. I would say it's, like, hard to identify. Hard to measure.}

Further, practitioners may recall areas with a high amount of TD but are sometimes unable or unwilling to articulate the problem constructively: \interviewQuote{I hear about \innerQuote{[vocable of complaint] this shitty part of the system.}}

These extracts tell us that software practitioners have trouble estimating and communicating the consequences of existing TD items. However, as suggested by one interviewee, they may hold strong intuitions. \interviewQuote{
In industry, it's more \innerQuote{I know something is wrong. It feels like things are spread out like this. I just can't put my finger on it.} \quoteSkip{} The feeling I have in industry is more like, \innerQuote{I know I'm gonna work in this area. I know it's gonna be horrible. I don't know what's gonna happen, exactly. Something is gonna show up. It's gonna take longer time. I can't give you a real estimate for how much it is to fix all of it, and I can't give you a real business value, because I just know it's gonna be hell.}}

In conclusion, the analysis reveals that affects are very much a key aspect of TD\@. They provide an insight into the underlying mechanics for how software practitioners respond to TD items. These \psychologicalRebounds{} may be a necessary consequence of TD and should not be ignored. The findings are further summarized in the following data extract and the box below (as findings F12--F24). \interviewQuote{
if it's [the debt is] manageable or if I feel I can fix it, then it feels a bit okay. It's like, \innerQuote{oh, this is a crappy thing someone did, but---whatever, it's fixable} in contrast to, like, \innerQuote{this is just a nest of---we just need to re-engineer.} That makes you just angry inside. \quoteSkip{} you can definitely feel when it's \innerQuote{[vocable of excitement], I can refactor this} or \quoteSkip{[vocable of quitting], this is such a mess. I hate going into this code. I can't fix a bug here, 'cause there's just going to pop up things in other places.} So, it's a mix. Depends on how much impact you can have on it, I think. Because it can be really fun to actually fix stuff. But when you can't, then it's like \innerQuote{[vocable of annoyance], angry.}} 

%the affects associated with TD, plus the behavioral responses, both appear to depend on the context in a non-trivial fashion. Polices and attitudes (individual, team, and company) towards TD management seem like a critical aspect, e.g., the prioritization of managing TD with regards to time pressure. Another important factor could be the technical safety nets, e.g., architecture and tests, and their ability to isolate subsystems and detect mistakes. 

\RQBox{
    \textbf{Findings for RQ3:}
    \begin{itemize}
        \item[F12] Software practitioners experience (strong) affects from TD along all three dimensions.
        \item[F13] When faced with high (overwhelming) levels of TD, practitioners will be reluctant to perform their work tasks.
        \item[F14] Time pressure is sometimes a catalyst for negative affects.
        \item[F15] Viewing TD items as opportunities for improvement appears to correlate with dominance toward the code base.
        \item[F16] TD anxiety relates to code dependencies, ripple effects, and (the risk of) defect introduction.
        \item[F17] TD anxiety appears to be correlated with submissiveness toward the code base.
        \item[F18] Displeasure plays an important role in recognizing the presence and severity of TD\@.
        \item[F19] Software practitioners sometimes get positive affects from amortizing TD\@.
        \item[F20] Profanity frequently emerges in TD discussions.
    \end{itemize}
    
    \bigskip
        
    \textbf{Additional findings:}
    \begin{itemize}
        \item[F21] Quality processes sometimes get disrupted by software practitioners' affects.
        \item[F22] Misalignment of quality expectations may result in interpersonal conflicts or burnout.
        \item[F23] TD is challenging to decode (recognize, estimate, and communicate).
        \item[F24] Violations of something the software practitioner considers fundamental appears to result in stronger affects.
    \end{itemize}
}

\section{Discussion \label{sec:discussion}} 
In this section, we tie together the results from the quantitative and qualitative analyses. We will continuously refer to the main findings (F1--F11 and F12--F24) at the end of Sections~\ref{sec:quantitative}--\ref{sec:qualitative}. First, we discuss the quantitative results related to the various scenarios and smells; to better explain the results, we then explore the quotes from the qualitative data occurring in correspondence with the analyzed smells. This allows us to explain how our results answer RQ1, or else how the smells influence the participants' affects. We compile a ranked list of which smells seem to have more impact. 

We also discuss how changes in affective state align with professional characteristics (RQ2). We then take a broader scope and reason on the exploratory results from the qualitative analysis and what relationships we have found between affective states and technical debt (RQ3). 

\subsection{Case A: Missing Encapsulation}

The quantitative analysis strongly suggests (F2) that the presence of the smell related to missing encapsulation in the code (ScA-H) causes the software practitioners to feel less pleasure (valence). This entails that practitioners consider the presence of this smell with disapproval rather than with indifference. 

We do not seem to find other significant evidence related to the other two dimensions (arousal and dominance), which could imply that the practitioners do not consider this smell exceedingly threatening. This is also mentioned in the qualitative data, as one of the participants mentioned: \interviewQuote{Like, some of them were quite [vocable of annoyance] as solutions, but didn't really impact me that much. Like, the rectangle whatever---PNG-things [reference to ScA-H]. Like, yeah, I can refactor this in an afternoon.}

On the other hand, such a lack of strong feelings could be caused by the limited size and localization of the example and how easy it is to estimate the practitioner's refactoring. One of the interviewees mentioned \interviewQuote{So, it's---this, like, rectangle-PNG-thing [reference to ScA-H]---it's, like, I can really point to it. Show it. I can give an estimate for how much time is left and how much impact it is. The feeling I have in industry is more like, \innerQuote{I know I'm gonna work in this area, I know it's gonna be horrible. I don't know what's gonna happen, exactly.}}

In conclusion, the smell is recognized as a problem, but not as a high-priority one. Suppose we consider the strength of the resulting feelings and the participants' insights for this smell. In that case, we can conclude that the presence of this smell, although frowned upon, is perhaps not considered detrimental by the software practitioners.

\subsection{Case B: Missing Hierarchy}

We did not find evidence to support the hypothesis that this smell generates any negative feeling in the software practitioners. Surprisingly, on the contrary, we found (moderate) evidence (F4) that practitioners felt more dominant (dominance) in working with the code containing the smell (ScB-H).

On the other hand, this scenario was mentioned a lot in the qualitative analysis in rather negative terms. However, those comments often referred to the whole code and not to the specific smell. Although, at the same time, some participants explicitly mentioned the smell and suggested the correct refactoring. \interviewQuote{Yeah, one example, that had the private class there [referring to ScB-H], and that one I didn't like \quoteSkip{} Yeah, overall the checking of types in code like that: I think it's a sign of bad architecture, most of the time, when you have to check the type of objects coming in. Then you can probably---yeah, like I said---interface it out. And an interface coming in and you have the method on the interface and, yeah.} 

The scenario itself could explain these seemingly contradictory results: debt-intense areas could bias the practitioner to distrust suitable constructs. After all, understanding and implementing a correct hierarchy involves a greater ability of abstraction. In other words, given that the original developers fell short in performing more straightforward tasks, confidence would be low to succeed in more demanding activities. As one of the interviewees said, in a different context, \interviewQuote{it's so, like, something I really think people should know how to do. It's so basic, in programming. So, yeah, I think so. It makes me a bit more worried, so to say, when I see that stuff. 'Cause it's very much easier to do correctly.}

Another, less plausible, explanation would be that practitioners feel submissive (intimidated) \emph{because} of the necessary abstraction skills, i.e., are not comfortable with such constructs. Here, it is essential to note the gap between recognizing a suspicious programming language construct (\texttt{instanceof}) and intimately understanding which abstraction would be suitable. The former is a low-level pattern detected by static code analysis (or even text search), while the latter often requires domain knowledge and experience. However, this seems unlikely, as most participants were experienced in object-oriented programming languages.

Finally, a third explanation could be that abstractions (by definition) remove details from the context. In other words, while beneficial for the system's maintainability, abstractions might, locally, result in less insight and, hence, less control.

In conclusion, the findings suggest that the smell is considered a problem despite its positive impact on dominance. The surrounding code's quality appears to confound individual TD items, but this effect needs to be verified in future studies.

\subsection{Case C: Broken Modularization }

Similar to Case B, we did not find evidence to support the hypothesis that this smell generates any negative feelings in the practitioners. However, we did find (strong evidence, F3) that they felt more pleasure (valence) and (moderate evidence) more in-control (dominance) when working with the code containing the smell (ScC-H).

This can be explained by the fact that the broken modularization smell consists of a widely recognized correct approach (modularization) applied in the wrong way (broken). In particular, the code that was modularized did not need to be (it consists of just variable declarations), and it should have been contained in the same abstraction (ScC-L). However, the participants' feelings might have been triggered by the presence of a better visual structure in ScC-H\@. The lack of a counter-effect for the displacement of the modularized code (ScC-H) could have different implications: 

\begin{ieeeEnumerate}
    \item The positive feelings in the presence of modularized code far outperforms the negative feelings related to the sub-optimal use of such mechanism. This is also supported by one of the participants: \interviewQuote{It's, like, I could sort of see what had happened, I think. Like, the last one [reference to ScC-L] with the weird device. It looked like a container of data and someone plonked helper methods in it, maybe, I don't know. It's, like, I can see how that happened. I can move them without changing anything, so there won't be any ripple effects and I can still improve the code, for instance.}
    \item The practitioners could have overlooked the specific code that was modularized in an additional class, focusing more on the structure rather than on the code itself. Alternatively, the participants might have thought that the additional class, the results of the modularization (which contains only variable declarations in our example), could have contained additional methods that were not displayed in our snippet. 
    \item While modularization is a well-known good practice, broken modularization is a less well-known bad practice among the practitioners. This can also be related to the language used by the participants and their familiarity with object-oriented programming. However, we did not find any evidence in the quantitative analysis supporting such an explanation.
\end{ieeeEnumerate}

In conclusion, we could not find evidence that this smell generates negative feelings in software practitioners. On the contrary, it seems as though the code with the smell was liked more, probably because the participants did not recognize (consciously or unconsciously) the misuse of modularization as significantly impacting.

\subsection{Case D: Cyclic Dependencies }

This is the smell for which we have quite strong evidence (F1) supporting hypotheses from literature \citep{martini2018identifying, al2014shape}. We can see how, for valence, the software practitioners reported extra-pleasure in the presence of code that is refactored (ScD-L), while at the same time, we register strong evidence that such code is much better liked than the one with the smell (ScD-H). Our analysis also reports moderate evidence for dominance (F6), where practitioners feel much more in control of refactored code (ScD-L) than the code containing the smell (ScD-H). 

Despite such strong results, the smell was not often or explicitly mentioned in the qualitative answers of the practitioners dealing with ScD-H, if not for the two quotes below, which can be related to this specific smell: 
\interviewQuote{The spontaneous feeling was a bit stress about too much stuff going on. Too many components. And some strange dependencies.}
and 
\interviewQuote{the code doesn't have to be perfect, or there could be some problems with the code. But if you have a great design, a great architecture, following the SOLID principles. That are loosely coupled. They are easy to fix.}
This could have happened because other smells or scenarios were more interesting to discuss, either because this example was not considered too challenging (perhaps because of the limited size of the example) or, possibly but perhaps less likely, because it was more noticeable and therefore less interesting.

In conclusion, we can consider this as evidence that the presence of cyclic dependencies generates stronger negative feelings in practitioners along at least two dimensions (valence and dominance).%, although developers did not find it interesting to discuss such issues. 

Although this can be somewhat expected (cyclic dependencies is a well-known smell, probably more than the other smells), it is interesting to note how the degree of negative feelings for this smell far exceeds other smells. We find this plausible: Cyclic dependencies is the smell that tends to involve multiple entities (usually classes), which can generate ripple effects across the code. Also, the example that we propose here consists of just one dependency. %However, the cyclic dependencies anti-pattern could consist of several dependencies and several involved entities, which would increase any harmful effect. Suppose one were to project the impact that we have recorded on the affective state in this simple example for a larger one. In that case, we could expect even stronger negative feelings affecting practitioners at the sight of this smell. On the other hand, this was a clear and obvious case of cyclic dependency. 
In contrast, dependencies, especially if involving several entities, can become less noticeable and not so visible if they are not explicitly investigated, as shown in other publications, see~\citet{martini2018identifying} and~\citet{al2014shape}. 

\subsection{Case E: Rebellious Hierarchy }

We did not find even moderate evidence that this smell would generate either positive or negative feelings in the participants concerning any of the dimensions. Therefore, it is difficult to draw conclusions on this smell and its impact on the practitioners' affects. In general, it seems as if the participants would be quite indifferent to one of the other proposed solutions. For example, even when encountering ScE-L, one of the practitioners mentioned:
\interviewQuote{you had the \texttt{\emph{Document}} [reference to ScE-L], yeah that was the wrong structure of the abstract class, I think, because you had all those methods, but only some of the implementation used. They didn't represent the same object. If you looked in the implementation, they had different actions or abilities. I think the public part of the implementations should be the same.} 

The lack of evidence in itself combined with the quotation could point to three possible conclusions: 

\begin{ieeeEnumerate}
    \item This smell is not considered a problem by practitioners, and it does not affect them.
    \item Our example was not a good representation of the actual issue. Unfortunately, we did not find an existing implementation that would suit our experiment, so we had to adapt our snippet from~\citet{refactoring-design-smells}, removing any domain-specific reference that our participants would not understand. This process might have excessively simplified the smell. 
    \item The smell consisted of one short method out of ten, distributed in four (ScE-H) and five (ScE-L) classes, respectively. This could mean that the participants might have overlooked it in the time allowed for the task, perhaps focusing their attention on other code features, such as its structure (as mentioned in the previous quotation from a participant).
\end{ieeeEnumerate}

In conclusion, we cannot draw many conclusions from these results, although we can speculate that the participants have not recognized this as an issue affecting their feelings.

\subsection{Comparison Across the Smells}

We have so far reported our reflections, based on available evidence, on how and why the different smells have impacted the participants' affects. However, can we say something more about how the different smells compare to each other? 

We report a summary (see \tabRef{tab:smell-prioritisation}), in which we compile a prioritized list of the smells based on the reported evidence. The smells are ordered by their negative impact on the software practitioners' affects. We also report if other impacts have been found on the refactored solution. To clarify the results, we have arranged the relationship positive\slash negative concerning the quantitative analysis, i.e., we do \emph{not} consider the direction of the SAM\@. For example, in \tabRef{tab:smell-prioritisation}, `negative impact on valence' means displeasure. We also highlight the strength and type of evidence supporting our conclusions.

\begin{table}
    \extrarowheight=\aboverulesep
    \addtolength{\extrarowheight}{\belowrulesep}
    \aboverulesep=0pt
    \belowrulesep=0pt
    \centering
    \caption{Prioritized smells according to our findings.}
        \begin{tabular}{@{}|p{2cm}|p{2.2cm}|p{1.5cm}|p{4.5cm}|@{}} \toprule
        \rowcolor[gray]{0.7}
        Smell & Impact on affects & Dimensions & Other considerations \\ \midrule
        \rowcolor{Red}
        D---Cyclic \newline dependencies & Negative \newline Negative (likely) & Valence \newline Dominance & Not explicitly discussed \newline qualitatively \\ \midrule
        \rowcolor{Orange}
        A---Missing \newline encapsulation & Negative & Valence & Easy to estimate a refactoring \\ \midrule
        \rowcolor{Orange}
        B---Missing \newline hierarchy & Positive (likely) & Dominance & Recognized as bad practice, but overshadowed by the scenario code \\ \midrule
        E---Rebellious \newline hierarchy & - & - & Seems to not have been recognized as an issue \\ \midrule
        \rowcolor{Green}
        C---Broken \newline modularization & Positive \newline Positive (likely) & Valence \newline Dominance & Modularization seems to give \newline positive feelings even if misused \\
        \bottomrule
    \end{tabular} 
    \label{tab:smell-prioritisation}
\end{table}

\subsection{Suggestions for Practitioners}

Based on \tabRef{tab:smell-prioritisation}, we can undoubtedly suggest practitioners pay attention, especially to cyclic dependencies and missing encapsulations. As for the latter one, the practitioners mention that its refactoring would not be costly, which could make it a good candidate for a mandatory cleanup of the code before release.

Less vigorously, we also suggest that practitioners keep their eyes out for missing hierarchies. While the presence of this smell \emph{increased} dominance that could be indicative of other problems, e.g., a need for domain knowledge acquisition, practitioners should consider taking action if developers start introducing this smell despite them recognizing it as bad practice.

As for the rebellious hierarchy, the study results do not allow us to draw firm conclusions, maybe because such smells might not be considered so upsetting by the participants. Finally, and probably surprising, it seems that \textit{a modularization that is not entirely correct is not considered problematic}. At the same time, developers might tend to consider the pleasure of modularized code (even if containing a smell) better than smell-free code, which might be less modularized. 

However, we need to notice that, for rebellious hierarchy and parts of broken modularization, the conclusions cannot be considered very strong, as our evidence is moderate. These results are also related to the influence of the smells on the participants' affects and do not consider other negative or positive effects. However, we consider our findings important to report, as developer unhappiness has been linked to harmful consequences (see \secRef{sec:TD-and-human-aspects}). Further, the results should not be confused with the actual extra-maintenance effect these smells have in practice, although the two variables are most probably correlated.

\subsection{The Effect of Experience \label{sec:effect-of-experience}}

Our quantitative results do not point to correlations between the participants' professional characteristics and how the affective states changed. The most striking results are related to the experience of the respondents.

Experience has shown (moderate evidence, F7) to have a negative impact on Dominance. In other words, the more work experience the subject had, the more submissiveness they report.

A possible explanation for the increased submissiveness is that more experienced practitioners have dealt with the technical debt related to the smells for a longer time than junior ones. This may be caused by the fact that they have witnessed more of the technical debt's long-term negative impact, which may trigger additional caution for the smells.

These results also seem in line with our qualitative findings, primarily related to the maturity (see next section) with which practitioners undergo TD: we could argue that, with less experience (and, probably, less maturity), practitioners seem to want to ignore TD and avoid to worry about it (as highlighted by the \procrastination{} theme), hence the presence of lower submissiveness. Then, moving to a more elitist and compensating attitude toward TD as they gain more experience, they become additionally worried when encountering TD (as shown in the feeling of lacking control, mentioned in relation to the \apprehension{} theme).

Also, these results are in line with what is reported concerning how startup teams are composed and their inclination to incur TD\@. \citet{besker2018embracing} report on interviewees from startups mentioning how it can be considered better to include a large part of junior developers in the initial team to make sure that TD is accrued (saving resources in the face of an initial high risk of failure). Experienced practitioners (apart from a small initial fraction) would be more suited for the growth and mature phase of a startup when TD needs to be removed before it becomes disruptive.

\subsection{The Overall Effect of TD on Affects}

Participants report that TD items activate a substantial portion of the emotional spectrum (three dimensions, F12), including vivid ones (e.g., profanity occurred, F20). Still, our experiment showed nothing concerning the arousal dimension. A plausible explanation is that participating in the experiment represents a different situation than encountering technical debt in real projects. The technical debt encountered during the experiment is not directly and negatively impacting the practitioners with, for example, extra-effort or additional bugs. This means that the arousal dimension could be triggered in a different context. 

Many participants receive satisfaction from improving code (F15, F19). Mainly, being able to perceive their work as impactful causes pleasure. On the contrary, the uncertainty caused by code affected by TD and the consequent distrust in the code base are sources of negative feelings (F13). Architectural TD is considered a common source of negative feelings, especially for problems related to ripple effects (as, for example, in case $D$ for cycling dependencies, F16).

Then the question is: why is TD so present in the software industry, and why is, e.g., code not continuously refactored?

First, as practitioners reported, stress is prevalent in the software industry. Several participants see deadlines as negatively affecting themselves and the product (F14). Avoiding TD requires more time, which would increase the stress in the presence of a deadline. This might mean that practitioners, to avoid stress, prefer to incur TD\@. Second, the participants mentioned that TD problems encountered in their daily lives are more extensive and more obscured than those in the experiment. %This most probably hinders them from being able to fix TD issues efficiently.

Another point of consideration was raised in the qualitative analysis, namely, that each sub-theme for undergoing TD is a \psychologicalRebound{}. Further, there seems to be a sort of progression to them, which we will refer to as \emph{maturity}, as we can draw parallels to our previous experience with group development models~\citep{grenTF17maturity}.

First, \procrastination{} (\simpleQuote{Forming}) can be interpreted as a mechanism with little interest in improving the situation. Consequently, the practitioner will not attempt to share the team's burden or attempt to shield its members from the harmful stimulus. Second, \elitism{} (\simpleQuote{Storming}) involves questioning the code base and the \emph{modus operandi}, which can be destructive and socially taxing unless adequately managed. Finally (note the absence of \simpleQuote{Norming}) \compensation{} (\simpleQuote{Performing}) illustrates a successful transition from defensive reactions to coping ones, with the participants focusing on facing up to the TD item and resolving it constructively.

\subsection{Comparison to Related Work}

In reviewing the literature, very little was found on the relationship between DTD and affective states, but several studies that investigated adjacent topics have found intriguing results. Many of the consequences of developer unhappiness demonstrated by~\citet{graziotin2018happens} were echoed in our qualitative findings, e.g., \textit{mental unease or disorder} (F13, F16, F17, F22). Perhaps most importantly, the quantitative analysis found a counterpart to \textit{lower code quality}: Some design smells elicited an unhappier response among the participants than did other such smells (F1, F2). This could indicate a vicious cycle where TD leads to more unhappiness, which in turn leads to more TD\@.

Our previous paper on the relationship between TD and morale,~\citet{morale-TD}, found that TD negatively impacts morale, but also that morale is increased by proper TD management. These results are corroborated by F13, F16, F17, and F22; and F15 and F19, respectively. Hence, we provide further evidence in favor of the long-held belief that morale and TD are intertwined \citep{exploration-TD, td-folklore}.

The findings in this paper also corroborate several of those of~\citet{lim2012balancing}, e.g., developers fearing certain parts of the code base (F12, F13, F16, F17) and TD being difficult to communicate (F20, F22, F23, F24).

\subsection{Implication for Research and Industry}

In this study, we conducted an empirical investigation that joined the fields of TD and PSE. As demonstrated by~\citet{graziotin2018happens}, affective states have important consequences for software engineering activities, and our findings provide solid evidence that design smells interlink with affective states. Accordingly, we present the argument that TD management should start factoring the human psyche into the decision-making processes.

\begin{figure}
    \centering
    \includegraphics[width=0.8\linewidth]{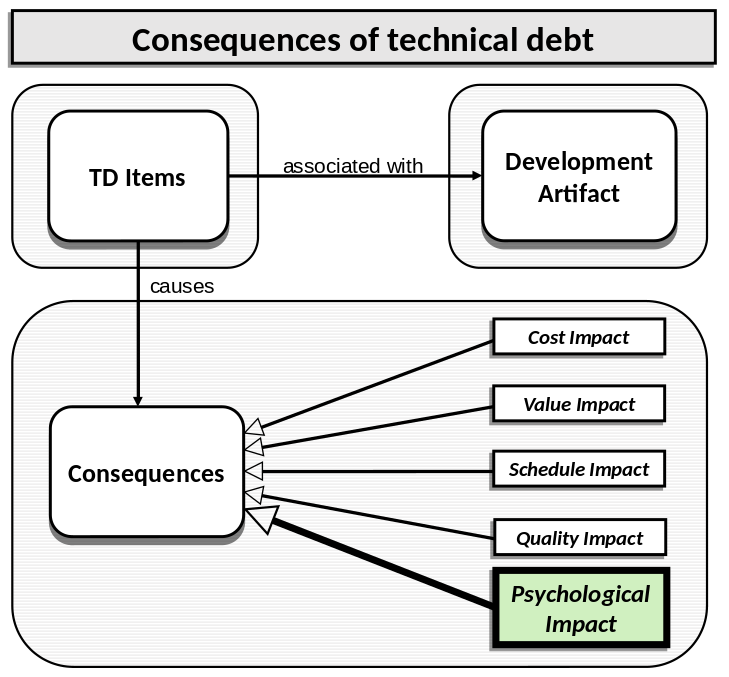}
    \caption{Partial view of a conceptual model of technical debt, adopted from~\citet{Dagstuhl}. The model has been extended to include the \textit{psychological impact} in \textit{consequences}.}
    \label{fig:extended-TD-map}
\end{figure}

Our findings, by themselves, constitute a compelling case, but do not stand alone. Although the human aspect is still a deficit area in the TD research, the combined results of~\citet{morale-TD, td-folklore, lim2012balancing, yli2014sources, exploration-TD}---many of which are corroborated by this paper---provide convincing grounds for our argument. Hence, we call for the research community to expand on the conceptual model of TD~\citep{Dagstuhl}. \figRef{fig:extended-TD-map} contains our proposition for how this part of the body of knowledge should be incorporated in our shared understanding of TD: The psychological factor would be explicitly acknowledged as a consequence of TD items. 

The reason for expanding the model is to nuance how the research community and the industry view TD\@. The fact that software engineering is a human activity is often overlooked in investigations. While we recognize the challenge in putting a value on such aspects, that does not mean that we can turn a blind eye to their actual costs and benefits. Recognizing the psychological factors in TD will serve as the starting point for discussions and help the community converge on key concepts.

In particular, we encourage software engineers in the industry to engage in introspection, especially concerning stress and burnout. As surfaced in the qualitative data, many professionals face a psychologically taxing work environment, and until the consequences of their experiences are better understood, we advise caution. From our own experiences, the digital work environment, partly constituent of the code base, is seldom (if at all) regarded in analyses of occupational safety, health, and welfare.

Another crucial goal of TD management is to prioritize the removal of debt items that generate the worst current and future negative effects (or else, to use the metaphor, they have high interest attached). As repeatedly reported in the literature, this is a very difficult task, as measuring such interest is challenging and evidence is scarce. Measuring the affective states of practitioners in relation to different TD issues can be used as a proxy for such interest, or can at least provide additional insights on which items are perceived as the most \simpleQuote{dangerous.} %In fact, negative effects have most probably be registered by practitioners' experience, which is then manifested via their feelings when they encounter TD (see \secRef{sec:effect-of-experience}). 
In \tabRef{tab:smell-prioritisation}, we provide a concrete example of how different smells (representing TD) impacted the participants' affective states differently, which suggests a ranking across the smells. In conclusion, a comprehensive catalog of smells and their impact on practitioners' affects could highly benefit the software engineering community. 

Finally, we would like to address the psychological perspective in the context of education. Software engineering is difficult work and perhaps especially so because of the flexibility of the medium. Unlike other constructs, a code base is largely unconstrained by natural forces and can thus be perpetuated to unfathomable complexity. Unsurprisingly, resignation, frustration and even hate ensue. Many universities may want to consider explicitly educating their students about this reality and train them in how to engineer under such conditions.\footnote{Project courses might \textit{not} be the optimal choice as the work situation likely differs from that in the industry, but more research is needed to determine this.} We argue that not imparting this knowledge would be an important oversight and urge the institutes to reflect on how our society is affected by the practitioner's emotional intelligence.

\section{Limitations and Threats to Validity}\label{sec:threats}

Conducting empirical studies with human subjects is often a complex issue~\citep{miller2008triangulation}, as it is often the case that the context is noisy and investigated effects are small~\citep{gelman18nhst}. As such, the potential threats to validity are often numerous, and it would be infeasible to discuss them all fully. This section presents what we consider the most significant validity threats to this study and the measures taken to mitigate them. The threats are categorized according to the aspects suggested by~\citet{wohlin2012experimentation} and~\citet{runeson2009guidelines}.

\subsection{Construct Validity}
This study set out with the aim of determining how DTD relates to the affects of software practitioners. One of the methods used was a repeated-measures experiment, where the participants were presented with five software design smells and their respective refactored versions. However, those smells were instantiated in code examples that originated from the same source, namely~\citep{refactoring-design-smells}, which is not a scientific publication (although a derivative of one). Also, the source's purpose differs from ours in that the smells and the refactored versions are intended to be contrasted with each other. As previously stated, the rationale of this choice was a perceived deficit of suitable DTD representations in the research literature (for our purposes, at least).

These characteristics introduce threats to validity, but because they were identified before the data collection, countermeasures could be introduced. Since we were unable to find a way to eliminate these threats, we chose to monitor them and investigate the issue \textit{post hoc}. This was done by introducing validity-checking questions (see \tabRef{tab:interview-questions}, questions IQ4.1, IQ4.2, and IQ5) to the interview questions and analyzing the answers. (Further, IQ3, IQ6, and IQ7 checked other types of validity.)

In response to IQ5, the participants reported that the scenarios were representative of industry code, albeit atypically small and isolated examples of TD encountered in practice. Additionally, they confirmed that industry code would have impacted their affects to a greater extent. This suggests that the treatment was suitable, but also that the resulting data is an \textit{underestimation} of the software industry's situation.

\subsection{Internal Validity}
The laboratory experiment part of this study was a repeated-measures design. While this approach lowers the threat of confounders because each subject's peculiarities are accounted for, it is more susceptible to learning effects.

Several countermeasures\footnote{Detailed in the replication package.} were taken to reduce learning effects. First, the participants were acquainted with the situation during the first phase of the session (pre-task instructions), and each received the same instructions for how to use the measurement instrument and their task. Second, the participants obtained practical experience with the procedure before the measurements (anchor point). Third, intermissions were used between measurements (deacclimatization periods) to lower the probability of any affects induced by previous scenarios carried over to later ones. Fourth, each participant was randomly allocated to one of two complementary treatment patterns designed to minimize bias. Fifth, and perhaps most importantly, \textit{the order of the scenarios} was randomized for each participant.

\subsection{External Validity}
For this study, it is worth noting that the field of psychology is experiencing a replication crisis. Unfortunately, we have been unable to find consensus on concrete best practices for ensuring replicability and have instead chosen to adopt some propositions.

We have made our work as transparent as possible (see the replication package), under the constraints set by confidentiality, anonymity, and copyright. This includes the statistical analysis, the data, the procedure, and the experiment material.

Another issue concerned with external validity is the sampling strategy. In this study, we employed convenience sampling (further detailed in the replication package). The approach meant that the sample was limited in several ways. First, all participants were industry professionals, which is a subset of all software practitioners and might not be representative. Second, the participants were selected by managers, who might have their agenda in what employees to select. Third, the companies belonged to the subset of companies that were both sufficiently interested in this study and could allocate resources (i.e., subjects).

However, our results show that the effects of different professional characteristics, such as programming language and role, were limited. This could indicate that the study is less susceptible to convenience sampling than otherwise. Further, the $40$ participants had a wide variety of professional backgrounds and were employed at twelve different companies and one government agency.

Along the same line, the generalizability of the results of the study is threatened by demographic factors. Due to various constraints (including financial), all partaking entities had offices in Sweden. While the study was conducted in several parts of the country, Sweden is culturally distinguished in terms of secular-rational and self-expression values~\citep{inglehart2010changing}. That said, there was diversity in, e.g., ethnicity among the participants, but such data was not collected to protect confidentiality. For the same reason, many aspects of the participants' demographic profiles were not investigated.

Finally, it is important to recognize that this study was an exploratory one, and not comprehensive. Hence, the quantitative findings should \textit{not} be understood as applying to all design smells nor all instances of the selected design smells. What the data demonstrates is that---even in the context of small, isolated code examples---software practitioners' affective states can change in the presence of certain design smells.

\subsection{Conclusion Validity}
As far as we can tell, no previous studies have investigated how DTD relates to affects. Consequently, the findings of this study cannot be compared and contrasted with the findings of others. Instead, they must be evaluated in isolation and are, therefore, more susceptible to incorrect inferences and conclusions.

Three triangulation techniques~\citep{miller2008triangulation} were adopted to combat these threats. First, the data were triangulated in the sense that the sessions were spread out over four weeks, and the participants were employed at different entities. Second, researcher triangulation was achieved as two researchers took part in all data gathering and interpretation. Third, methodological triangulation was used, as data were collected through an experiment, a questionnaire, and interviews.

\section{Conclusion}\label{sec:conclusion}

Fully understanding the impact of technical debt (TD) in the code base is a crucial challenge for researchers and practitioners alike. Although previous research has highlighted how TD can impact developers' morale, there is scarce evidence on how specific technical debt issues impact practitioners' affective states. Even more challenging is finding evidence related to design and architectural debt.

With our study, encompassing a quantitative data collection and analysis supported by additional qualitative insights from the participants, we offer a first detailed look into how the presence of design debt issues affect software practitioners' affective state.  

The results show that five different smells have different impacts. Even when present in a small example, cyclic dependencies clearly and negatively affect software practitioners' affects. Simultaneously, missing encapsulation seems to be a more straightforward issue to deal with (although mildly affecting the practitioners' affects). Two issues related to hierarchy (missing hierarchy and rebellious hierarchy) seem to have a conflicting or no evident effect on the participants' affective state. In contrast, surprisingly, the presence of the broken modularization issue seems to have a positive impact on practitioners' affects. 

These results imply that these different TDs need to be treated differently and that studying their impact on the practitioners' affective states helps to understand their overall impact (interest) and consequently how to prioritize them in practice. More studies with additional TD should be studied in a similar way as it was done in this study, so to provide a comprehensive catalog of the smells and their impact.

From our qualitative findings, it seems that practitioners undergo different levels of maturity in how they deal with TD\@. First, they might naively tend to avoid it (\procrastination{}), then they tend to build a quality-heavy mindset (mostly, however, by blaming others for the presence of TD, i.e., \elitism{}). Finally, they reach a higher level of maturity when a constructive mindset promotes high-quality code (\compensation{}). Also, practitioners seem to be affected negatively when they forecast TD, especially with \apprehension{} related to the future negative impact generated by TD, and by the inherent difficulty in identifying TD and predicting its consequences (TD as \indeterminable{} items).

Finally, we investigated whether participants' background covariates played a role, and we found partly how experience seems to act as a sort of amplifier for the participants' feelings, probably due to repeated encounters with TD and to the different maturity, acquired with more experience, in dealing with TD\@.

In summary, only some of the known issues highlighted in the literature seem to affect practitioners' feelings. At the same time, we find that dealing with TD is stressful and might require a fair amount of experience in the team to be handled constructively.

This topic remains mostly uncharted, and presents many opportunities for future work. A singular study is insufficient to build a solid theory, but we encourage others to replicate our experiment under similar or different settings, e.g., design smells, TD type, or cultures. Two particularly interesting investigations would be using industry code examples and situations that simulate time pressure.

%Similarly, we discovered several peripheral and related research topics. Investigations of burnout, concerning TD and release deadline (or, in the case of continuous delivery, lack thereof), would be most welcome. As would further studies on the idea of psychological \emph{maturity} towards TD\@.

% \section{Conclusion \label{sec:conclusion}}
%\input{}

\begin{acknowledgements}
We want to thank all the participating companies, the individual participants in our study, and all the students who participated in our pilot studies. 

The computations were enabled by resources provided by the Swedish National Infrastructure for Computing (SNIC), partially funded by the Swedish Research Council through grant agreement no.\ 2018--05973.

\end{acknowledgements}

\bibliographystyle{spbasic} 
\bibliography{References} % Entries are in the "refs.bib" file

\end{document}